\documentstyle[epsfig,12pt]{article}

\newcommand{\be}{\begin{equation}}
\newcommand{\ee}{\end{equation}}
\newcommand{\ba}{\begin{eqnarray}}
\newcommand{\ea}{\end{eqnarray}}
\newcommand{\bd}{\begin{displaymath}}
\newcommand{\ed}{\end{displaymath}}

\def\thalf{{\textstyle{\frac{1}{2}}}}

\def\oneth{{\textstyle{\frac{1}{3}}}}
\def\twoth{{\textstyle{\frac{2}{3}}}}

\title{
{\bf Quasi-Particle Theory of Shear and Bulk Viscosities of Hadronic Matter}}
\author{{P. Chakraborty and J. I. Kapusta } \vspace*{0.1in}\\
{\it School of Physics and Astronomy, University of Minnesota}\\
 {\it Minneapolis, Minnesota 55455, USA}}

\date{}
\begin{document}

\maketitle

\begin{abstract}
A theoretical framework for the calculation of shear and bulk viscosities of 
hadronic matter at finite temperature is presented.  The framework is based on 
the quasi-particle picture.  It allows for an arbitrary number of hadron species 
with point-like interactions, and allows for both elastic and inelastic 
collisions.  Detailed balance is ensured.  The particles have temperature 
dependent masses arising from mean field or potential effects, which maintains 
self-consistency between the equation of state and the transport coefficients.  
As an example, we calculate the shear and bulk viscosity in the linear $\sigma$ 
model. The ratio of shear viscosity to entropy density shows a minimum in the 
vicinity of a rapid crossover transition, while the ratio of bulk viscosity to 
entropy density shows a maximum.
\end{abstract}

\vspace{0.5cm} 
\noindent {PACS numbers: 11.10.Wx, 25.75.Nq, 51.20.+d, 25.75.-q}

\vspace{0.2cm}
\noindent Keywords: Shear viscosity, bulk viscosity, hadronic matter at finite temperature, quark-gluon plasma, linear sigma model.

\newpage

\section{Introduction}

One of the amazing {\it experimental} discoveries of measurements of heavy ion 
collisions at the Relativistic Heavy Ion Collider (RHIC) is the surprising 
amount of collective flow exhibited by the outgoing hadrons.  Collective flow is 
observed in both the single-particle transverse momentum distribution 
\cite{radial} (radial flow) and in the asymmetric azimuthal distribution around 
the beam axis \cite{RHICv2} (elliptic flow).  It is now generally accepted that 
collective flow is mostly generated early in the nucleus-nucleus collision and 
is present before partons fragment or coalesce into hadrons \cite{coalesce}.  
The quark-gluon matter created in these collisions must be strongly interacting, 
unlike the type of weakly interacting quark-gluon plasma expected to occur at 
very high temperatures on the basis of asymptotic freedom \cite{pQCD}.  Perfect 
fluid dynamics with zero viscosity reproduces the measurements of radial and 
elliptic flow quite well up to transverse momenta on the order 1.5 GeV/c 
\cite{hydro}.  These results have been interpreted as strong indicators of early 
thermalization and collective flow on a time scale of several fm/c.

An amazing {\it theoretical} discovery was made by Kovtun, Son and Starinets 
\cite{Kovtun:2004de}.  They showed that certain special field theories (AdS/CFT 
or Anti-deSitter/Conformal Field Theory) that are dual to black branes in higher 
space-time dimensions \cite{Mald}-\cite{Polyakov} have the ratio of shear 
viscosity to entropy density $\eta/s = 1/4\pi$ (in natural units with $\hbar = 
k_{\rm B} = c =1$).  The connection between transport coefficients and gravity 
arises because both involve commutators of the stress-energy-momentum tensor.  
They conjectured that all substances have this value as a lower limit, and gave 
as examples various atomic and molecular systems.  In fact, it had been argued 
much earlier that any substance should a lower bound on $\eta/s$ because of the 
uncertainty principle \cite{Danielewicz:1984ww}.  Is the RHIC data telling us 
that the created matter has a very small viscosity, the minimal value of 
$\eta/s$, that it is a 
{\it perfect fluid}?

The relatively good agreement between perfect fluid calculations and 
experimental data for hadrons of low to medium transverse momentum at RHIC 
suggests that the viscosity is small.  However, it cannot be zero.  Indeed, 
calculations within AdS/CFT suggest that $\eta \ge s/4\pi$ 
\cite{stretched}-\cite{Springer2}.  (Whether or not this is a rigorous lower 
bound is still an open question \cite{Cohen:2007qr}-\cite{Jakovac:2009xn}.)  
There are strong theoretical arguments, and evidence from atomic and molecular 
systems, that $\eta/s$ should be a minimum in the vicinity of the phase 
transition or rapid crossover between hadronic matter and quark-gluon plasma 
\cite{Csernai:2006zz,singular}, and that the ratio of bulk viscosity to entropy 
density $\zeta/s$ should be a maximum there \cite{Kapusta:2008vb}.  See also 
\cite{Kharzeev1}-\cite{PaulSon}.  

It ought to be possible to extract numerical values of the viscosities in heavy 
ion collisions via scaling violations to perfect fluid flow predictions 
\cite{Bonasera}-\cite{Lacey:2006bc}.  The program is to solve relativistic 
viscous fluid equations, with appropriate initial conditions and with a hadron 
cascade afterburner \cite{burner}, over a range of beam energies and nuclei and 
extract $\eta(T)/s(T)$ and $\zeta(T)/s(T)$ from comparison with data.  Thus, 
sufficiently precise calculations and measurements should allow for a 
determination of the ratio $\eta/s$ as well as the ratio of bulk viscosity to 
entropy density $\zeta/s$ as functions of temperature, and that these ratios can 
pinpoint the location of the phase transition or rapid crossover from hadronic 
to quark and gluon matter.  This is a different method than trying to infer the 
equation of state of QCD in the form of pressure $P$ as a function of 
temperature $T$ or energy density $\epsilon$.  Because of advances in both 
theory and computation, vigorous activities are currently underway to determine 
the dissipative effects in heavy ion collisions
\cite{Luzum}-\cite{Rajagopal:2009yw}.

From the theoretical perspective, it should be possible to compute the shear and 
bulk viscosities directly from QCD at finite temperature.  In practice, this is 
extremely difficult because QCD is generally a strongly interacting theory.  
Calculations can and have been done at extremely high temperatures where 
perturbation theory, applied to quarks and gluons, can be used on account of 
asymptotic freedom; see \cite{AMY} for shear viscosity and \cite{Arnold:2006fz} 
for bulk viscosity.  At extremely low temperatures, perturbation theory can 
again be used because the matter consists only of a very dilute gas of pions, 
and low energy pion dynamics is well understood.  See \cite{Prakash:1993bt} for 
massive pions; for massless pions, see \cite{Prakash:1993bt} for shear viscosity 
and \cite{Chen:2007kx} for bulk viscosity.  There have been a variety of other 
kinetic theory calculations of the shear and/or bulk viscosities at low to 
moderate temperatures in the literature in recent years 
\cite{MurongaTransport}-\cite{Sasaki}; these usually include only elastic 
scattering of one or a few species of hadrons.  

In the intermediate region, which may be loosely defined as $100 < T < 400$ MeV, 
neither the low nor high temperature approach is accurate.  A few lattice QCD 
simulations have used the Kubo formulae \cite{Kubo} to compute the shear 
\cite{Nakamura:2004sy,MeyerShear} and bulk \cite{MeyerBulk} viscosities just 
above the critical temperature of pure gluon/glueball matter.  However, accurate 
lattice QCD simulations of the properties of hadronic matter are extremely time 
consuming and the final results are still likely to be far in the future.  The 
reason is that the lattice spacing $a$ must be small enough to describe the 
properties of an individual hadron ($a < 0.05$ fm) while the box size $L$ must 
be large enough to contain many hadrons forming the dilute gas ($L > 10$ fm).  
Hence the number of spatial lattice sites should be at least 200 in each 
direction.  An interesting alternative approach to the intermediate region is a 
model of classical, non-relativistic quasi-particles with color charges 
\cite{cQCD}. 

Our goal in this paper is to provide a theoretical framework in which to 
calculate the shear and bulk viscosities of hadronic matter.  This framework has 
the following features.
\begin{enumerate}
\item
It is relativistic.
\item
It allows for an arbitrary number of hadron species.
\item
It allows for both elastic and inelastic collisions.
\item
It respects detailed balance.
\item
It allows for mean fields and temperature-dependent masses.
\item
The viscosities and the equation of state are mutually consistent in the sense 
that the same interactions are used to compute them all.
\end{enumerate}
Obviously some assumptions or approximations must be made for the theory to be 
applied in practice.  The essential assumptions are that quasi-particles are 
well-defined and that the elementary interactions are local.  Thus our proposed 
theoretical framework goes well beyond the classic works of \cite{degroot} and 
\cite{Prakash:1993bt} which, although relativistic, considered only elastic 
collisions in dilute gases.  The inclusion of not only resonances, but 
especially inelastic collisions, mean fields, and temperature-dependent masses 
are essential for an accurate determination of the bulk viscosity \cite{Paech}.

The outline of this paper is as follows.  In section 2 we recall the basics of 
the Boltzmann transport equation.  In section 3 we derive the integral equations 
for the viscosities by using the Boltzmann equation.  In section 4 we show how 
the Landau-Lifshitz condition plays a crucial role for the bulk viscosity.  In 
section 5 we work out formulas for the viscosities in the relaxation time 
approximation.  In section 6 we generalize the previous results to include mean 
field or potential effects and their significance for the bulk viscosity.  In 
section 7 we apply the framework to the linear $\sigma$ model with massive pions.  
As expected, the ratio $\eta/s$ has a minimum and the ratio $\zeta/s$ has a 
maximum near the rapid crossover transition, which is more pronounced for larger 
vacuum $\sigma$ masses.  We conclude in section 8.  The reader not interested in 
mathematical details is referred to sections 7 and 8 and to the appendix where 
the main formulas are summarized.   

Since the asymmetry between matter and anti-matter in high energy nuclear 
collisions at RHIC is very small, so are the baryon and electric charge chemical 
potentials.  Therefore, thermal and electrical conductivity are neglected in 
this paper.  There inclusion is straightforward but tedious, and work to include 
them is in progress.

For ease and clarity of presentation we will, for the most part, display 
formulas that include only $2 \rightarrow 2$ reactions, both elastic and 
inelastic, and formation $2 \rightarrow 1$ and decay $1 \rightarrow 2$ of 
resonances.  At certain key points in the paper we simply write down formulas 
for the more general cases.  In addition, in practice we generally use classical 
statistics.  The lightest hadron for which this would be the most significant 
approximation is the pion, but even then the difference between Bose-Einstein 
and classical statistics have been shown to be inconsequential for the transport 
coefficients for zero chemical potential \cite{Davesne:1995ms}.  While some of 
the material here is not original, it is basic to developing the theoretical 
framework.  We present it to make the paper self-contained and to define our 
notation. 

\section{Boltzmann Equation}

The rate for $2 \rightarrow 2$ processes, that is, the number of reactions per 
unit time per unit volume of the type $a+b \rightarrow c+d$ is
\bd
{\rm rate} = \frac{1}{1 + \delta_{ab}}
\int \frac{d^3p_a}{2E_a(2\pi)^3} \frac{d^3p_b}{2E_b(2\pi)^3}
\frac{d^3p_c}{2E_c(2\pi)^3} \frac{d^3p_d}{2E_d(2\pi)^3}
|{\cal M}(a,b|c,d)|^2
\ed
\be
\times (2\pi)^4 \delta^4\left( p_a+p_b-p_c-p_d\right) 
 f_a f_b \left( 1 + (-1)^{2s_c} f_c \right) \left( 1 + (-1)^{2s_d} f_d \right)
\label{rate2to2}
\ee
\bd
= \frac{1}{1 + \delta_{ab}} 
\int \frac{d^3p_a}{(2\pi)^3} \frac{d^3p_b}{(2\pi)^3}
\frac{d^3p_c}{(2\pi)^3} \frac{d^3p_d}{(2\pi)^3} W(a,b|c,d) 
\ed
\be
\times f_a f_b
\left( 1 + (-1)^{2s_c} f_c \right) \left( 1 + (-1)^{2s_d} f_d \right)
\ee
whence
\be
W(a,b|c,d) = \frac{(2\pi)^4 \delta^4\left( p_a+p_b-p_c-p_d\right)}
{2E_a2E_b2E_c2E_d} |{\cal M}(a,b|c,d)|^2 \, .
\ee
Here $s_a$ is the spin of particle $a$, etc. and the factor $1/(1+\delta_{ab})$ 
takes into account the possibility that the incoming particles are identical.  
The amplitude ${\cal M}$ is dimensionless.  There are either Bose-enhancment or 
Pauli-suppression factors in the final state.  In the rest frame of the system 
the single-particle distributions $f_a$ are normalized such that
\be
\int \frac{d^3p}{(2\pi)^3} f_a({\bf x},{\bf p},t) = n_a({\bf x},t)
\ee
is the spatial density of particles of type $a$.  In thermal equilibrium
\be
f_a({\bf x},{\bf p},t) = \frac{1}{{\rm e}^{(E_a-\mu_a)/T} - (-1)^{2s_a}}
\ee
where $T$ is the temperature and $\mu_a$ is the chemical potential of the 
particle.  It is obvious that the rate is a Lorentz scalar.

The rate for decay processes, that is, the number of decays per unit time per 
unit volume of the type $a \rightarrow c+d$ is
\bd
{\rm rate} = \int \frac{d^3p_a}{2E_a(2\pi)^3} 
\frac{d^3p_c}{2E_c(2\pi)^3} \frac{d^3p_d}{2E_d(2\pi)^3}
|{\cal M}(a|c,d)|^2
\ed
\bd
\times (2\pi)^4 \delta^4\left( p_a-p_c-p_d\right) 
 f_a \left( 1 + (-1)^{2s_c} f_c \right) \left( 1 + (-1)^{2s_d} f_d \right)
\ed
\be
= \int \frac{d^3p_a}{(2\pi)^3} 
\frac{d^3p_c}{(2\pi)^3} \frac{d^3p_d}{(2\pi)^3} W(a|c,d) 
f_a \left( 1 + (-1)^{2s_c} f_c \right) \left( 1 + (-1)^{2s_d} f_d \right)
\label{rate1to2}
\ee
whence
\be
W(a|c,d) = \frac{(2\pi)^4 \delta^4\left( p_a-p_c-p_d\right)}
{2E_a2E_c2E_d} |{\cal M}(a|c,d)|^2 \, .
\ee
This ${\cal M}$ has dimension of energy.

How do the $W's$ relate to cross-sections and decay rates?  The relationships 
for the cross-sections are as follows.
\ba
\frac{d\sigma}{d\Omega^*} &=&
\frac{1}{64\pi^2 s} \frac{p^*_{\rm final}}{p^*_{\rm initial}}
|{\cal M}|^2 \nonumber \\
\frac{d\sigma}{dt} &=&
\frac{1}{64\pi s} \frac{1}{(p^*_{\rm initial})^2}
|{\cal M}|^2
\label{xsection}
\ea
Here $s$ and $t$ are the Mandelstam variables (one can tell from the context 
whether $t$ represents a Mandelstam variable or time).  The differential 
cross-section in the center-of-momentum frame is $d\sigma/d\Omega^*$ while 
$d\sigma/dt$ is usually written as a function of the invariants $s,t,u$.  Thus
\be
W(a,b|c,d) = \frac{s}{E_aE_bE_cE_d}
\frac{p^*_{\rm initial}}{p^*_{\rm final}}
\frac{d\sigma}{d\Omega^*}
(2\pi)^6 \delta^4\left( p_a+p_b-p_c-p_d\right) \, .
\ee
This agrees with the literature on relativistic Boltzmann equations when only 
elastic collisions are considered since then $p^*_{\rm final}= p^*_{\rm 
initial}$.  An alternative form, which may be more useful for cross-sections 
that are not isotropic in the center-of-momentum frame, is
\be
W(a,b|c,d) = \frac{2s(p^*_{\rm initial})^2}{E_aE_bE_cE_d} \frac{d\sigma}{dt}
(2\pi)^5 \delta^4\left( p_a+p_b-p_c-p_d\right)
\ee
where
\ba
4s\left(p^*_{\rm initial}\right)^2 &=& (s-m_a^2-m_b^2)^2-4m_a^2m_b^2 \, , \nonumber \\
4s\left(p^*_{\rm final}\right)^2 &=& (s-m_c^2-m_d^2)^2-4m_c^2m_d^2 \,.
\ea

For the decay $a \rightarrow c+d$ consider the particle $a$ at rest.  It will 
decay according to the usual exponential law.
\be
\frac{dn_a(t)}{dt} = -\Gamma_{a \rightarrow c+d} \, n_a(t) \,.
\ee
One computes that
\be
\Gamma_{a \rightarrow c+d} = \frac{p^*_{\rm final}}{8\pi m_a^2}
|{\cal M}(a|c,d)|^2
\label{Gammaformula}
\ee
where
\be
4m_a^2 \left(p^*_{\rm final}\right)^2 = 
(m_a^2-m_c^2-m_d^2)^2-4m_c^2m_d^2
\ee
so that
\be
W(a|c,d) = \frac{\pi m_a^2}{E_aE_cE_d p^*_{\rm final}}
\Gamma_{a \rightarrow c+d}  
(2\pi)^4 \delta^4\left( p_a-p_c-p_d\right) \, .
\ee

Now we consider the Boltzmann equation.  Taking into account both gain and loss 
rates we write it as follows.
\bd
\frac{\partial f_a}{\partial t} + {\bf v}_a \cdot \nabla f_a =
\sum_{bcd} \int \frac{d^3p_b}{(2\pi)^3} 
\frac{d^3p_c}{(2\pi)^3} \frac{d^3p_d}{(2\pi)^3}
\ed
\bd 
\times \Bigg\{ \frac{1}{1 + \delta_{cd}} W(c,d|a,b) 
f_c f_d \left( 1 + (-1)^{2s_a} f_a \right) \left( 1 + (-1)^{2s_b} f_b \right)
\ed
\bd
- \frac{1}{1 + \delta_{ab}} W(a,b|c,d) 
f_a f_b \left( 1 + (-1)^{2s_c} f_c \right) \left( 1 + (-1)^{2s_d} f_d \right) 
\Bigg\}
\ed
\bd
+ \sum_{cd} \int \frac{d^3p_c}{(2\pi)^3} \frac{d^3p_d}{(2\pi)^3}
\Bigg\{ \frac{1}{1 + \delta_{cd}} W(c,d|a)
f_c f_d \left( 1 + (-1)^{2s_a} f_a \right)  
\ed
\bd
 - W(a|c,d) f_a
\left( 1 + (-1)^{2s_c} f_c \right) \left( 1 + (-1)^{2s_d} f_d \right)
\Bigg\}
\ed
\bd
+ \sum_{bc} \int \frac{d^3p_b}{(2\pi)^3} \frac{d^3p_c}{(2\pi)^3}
\Bigg\{ W(c|a,b) f_c
\left( 1 + (-1)^{2s_a} f_a \right) \left( 1 + (-1)^{2s_b} f_b \right)  
\ed
\be
 - \frac{1}{1 + \delta_{ab}} W(a,b|c)
f_a f_b \left( 1 + (-1)^{2s_c} f_c \right) 
\Bigg\}
\label{Boltz1}
\ee

Due to detailed balance on the microscopic level, energy conservation, and 
chemical equilibrium as represented by $\mu_a + \mu_b = \mu_c + \mu_d$ for 2-
body reactions and by $\mu_a = \mu_c + \mu_d$ for 2-body decays, we find 
relations between the forward and backward going rates.
\ba
\left( 1+\delta_{ab} \right) W(c,d|a,b) &=& 
\left( 1+\delta_{cd} \right) W(a,b|c,d) \\
W(c,d|a) &=& \left( 1+\delta_{cd} \right) W(a|c,d)
\ea
The Boltzmann equation then becomes
\bd
\frac{\partial f_a}{\partial t} + {\bf v}_a \cdot \nabla f_a =
\sum_{bcd} \frac{1}{1 + \delta_{ab}} \int \frac{d^3p_b}{(2\pi)^3} 
\frac{d^3p_c}{(2\pi)^3} \frac{d^3p_d}{(2\pi)^3} W(a,b|c,d)
\ed
\bd 
\times \Bigg\{ 
f_c f_d \left( 1 + (-1)^{2s_a} f_a \right) \left( 1 + (-1)^{2s_b} f_b \right)
\ed
\bd
- f_a f_b \left( 1 + (-1)^{2s_c} f_c \right) \left( 1 + (-1)^{2s_d} f_d \right) 
\Bigg\}
\ed
\bd
+ \sum_{cd} \int \frac{d^3p_c}{(2\pi)^3} \frac{d^3p_d}{(2\pi)^3} W(a|c,d) 
\Bigg\{ f_c f_d \left( 1 + (-1)^{2s_a} f_a \right)  
\ed
\bd
 - f_a \left( 1 + (-1)^{2s_c} f_c \right) \left( 1 + (-1)^{2s_d} f_d \right)
\Bigg\}
\ed
\bd
+ \sum_{bc} \int \frac{d^3p_b}{(2\pi)^3} \frac{d^3p_c}{(2\pi)^3}
W(c|a,b) \Bigg\{ f_c
\left( 1 + (-1)^{2s_a} f_a \right) \left( 1 + (-1)^{2s_b} f_b \right)  
\ed
\be
 - f_a f_b \left( 1 + (-1)^{2s_c} f_c \right) 
\Bigg\} \, .
\ee

For classical statistics, where the Bose and Pauli factors are dropped, the 
equilibrium phase space distribution is
\be
f_a^{\rm eq}({\bf x},{\bf p},t) = {\rm e}^{-(E_a-\mu_a)/T} \, .
\ee
The Boltzmann equation then shortens somewhat.
\bd
\frac{\partial f_a}{\partial t} + {\bf v}_a \cdot \nabla f_a =
\sum_{bcd} \frac{1}{1 + \delta_{ab}} \int \frac{d^3p_b}{(2\pi)^3} 
\frac{d^3p_c}{(2\pi)^3} \frac{d^3p_d}{(2\pi)^3} W(a,b|c,d)
\left\{ 
f_c f_d - f_a f_b \right\}
\ed
\bd
+ \sum_{cd} \int \frac{d^3p_c}{(2\pi)^3} \frac{d^3p_d}{(2\pi)^3} W(a|c,d) 
\left \{ f_c f_d - f_a \right\}
\ed
\be
+ \sum_{bc} \int \frac{d^3p_b}{(2\pi)^3} \frac{d^3p_c}{(2\pi)^3} W(c|a,b) 
\left \{ f_c - f_a f_b \right\}
\label{Boltzeq}
\ee 

The generalization to arbitrary reactions $\{i\} \rightarrow \{j\}$ with $n$ particles in the initial state and $m$ particles in the final state is now clear.  In obvious notation the Boltzmann equation is
\be
\frac{\partial f_a}{\partial t} + {\bf v}_a \cdot \nabla f_a =
\sum_{\{i\}\{j\}} \frac{1}{S} \int^{\prime} dP_i \, dP_j
W(\{i\}|\{j\}) F\left[f\right] \, ,
\label{boltz_gen}
\ee
where the prime indicates that there is no integration over the momentum of $a$.  
There is a statistical factor for identical particles in the initial state
\be
S = \prod_i n_i!
\ee
and products of Bose-Einstein and Fermi-Dirac distributions as appropriate
\be  
F\left[ f \right] = \prod_i \prod_j \left\{ f_j  
\left(1 + \left(-1\right)^{s_i}f_i\right) -  f_i
\left(1 + \left(-1\right)^{s_j}f_j\right) \right\} \,.
\ee

\section{Viscosities}

Now we use the Boltzmann equation to calculate the viscosities.  We restrict 
ourselves to zero chemical potentials.  We assume that the system is in 
approximately local equilibrium, with local temperature $T(x)$ and flow velocity 
$U^{\mu}(x)$.  In the Landau-Lifshitz approach, $U^{\mu}(x)$ is the velocity of 
energy transport while in the Eckart approach, $U^{\mu}(x)$ would be the 
velocity of baryon number flow \cite{fluid1,fluid2}.  However, the net baryon 
number, electric charge, and all other conserved quantum numbers are taken to be 
zero.  Therefore one cannot use the Eckart approach.  Another consequence of all 
conserved quantum numbers being zero is that thermal conductivity has no meaning.

The symmetric energy-momentum tensor is written as
\be
T^{\mu\nu} = -Pg^{\mu\nu} + w U^{\mu} U^{\nu} + \Delta T^{\mu\nu}
\ee
where $P=P(T)$ is pressure, $s=dP/dT$ is entropy density, $\epsilon = -P+Ts$ is 
energy density, and $w=Ts=P+\epsilon$ is enthalpy density.  These are all 
measured in a frame in which the fluid is instantaneously at rest.  The 
$\Delta T^{\mu\nu}$ is the dissipative part.  It satisfies the condition
\be
U_{\mu} \Delta T^{\mu\nu} = 0
\label{LLdef}
\ee
on account of the Landau-Lifshitz definition of flow. The entropy current is
\be
s^{\mu} = s U^{\mu}
\ee
and is conserved if dissipative terms are neglected.  The most general form of 
$\Delta T^{\mu\nu}$ is given by
\be
\Delta T^{\mu\nu} =
\eta \left( D^{\mu} U^{\nu} + D^{\nu} U^{\mu} 
+ \twoth \Delta^{\mu\nu} \partial_{\rho} U^{\rho} \right)
 - \zeta \Delta^{\mu\nu} \partial_{\rho} U^{\rho} \, .
\label{Tdiss}
\ee
Here
\be
\Delta^{\mu\nu} = U^{\mu} U^{\nu} - g^{\mu\nu}
\ee
is a projection tensor normal to $U^{\mu}$, and
\be
D_{\mu} = \partial_{\mu} - U_{\mu} U^{\beta} \partial_{\beta}
\ee
is a derivative normal to $U^{\mu}$. The $\eta$ is the shear viscosity and the 
$\zeta$ is the bulk viscosity.  In the local rest frame of the fluid
\ba
\Delta^{0\nu} &=& 0 \nonumber \\
\Delta^{ij} &=& \delta^{ij}
\ea
and
\ba
D_0 &=& 0 \nonumber \\
D_i &=& \partial_i \, .
\ea
In this frame 
\be
\partial_{\mu}s^{\mu} = \frac{\eta}{2T}
\left( \partial_iU^j + \partial_jU^i - \twoth \delta^{ij} \nabla \cdot {\bf U}
\right)^2 + \frac{\zeta}{T} \left( \nabla \cdot {\bf U} \right)^2 \, .
\ee
Non-decrease of entropy requires that both viscosities be non-negative.
   
We are assuming that the interactions are well localized in space and time.  We 
are assuming that they are, for practical purposes, point or contact 
interactions.  Then the energy-momentum tensor is written as a sum of 
independent contributions.
\be
T^{\mu\nu}(x) = \sum_a \int \frac{d^3p}{(2\pi)^3} \frac{p_a^{\mu} p_a^{\nu}}
{E_a} f_a(x,p)
\ee
Allow the system to be slightly out of equilibrium.  This means that 
$U^{\mu}(x)$ is not constant in space and time, but that departures from local 
equilibrium are small.  Then we can write
\be
f_a(x,p) = f_a^{\rm eq}(U_{\alpha} p^{\alpha}/T)
 \left[ 1 + \phi_a(x,p) \right]
\label{departure}
\ee
and so
\be
\Delta T^{\mu\nu} = \sum_a \int \frac{d^3p}{(2\pi)^3} \frac{p_a^{\mu} p_a^{\nu}}
{E_a} f_a^{\rm eq}(U_{\alpha} p^{\alpha}/T) \phi_a(x,p)
\label{deltaTuv}
\ee
where $|\phi_a| \ll 1$.  There is a constraint on $\phi_a(x,p)$ in that the 
Landau-Lifshitz condition (\ref{LLdef}) must be satisfied.  It is customary and 
natural to use the same tensorial decomposition of (\ref{Tdiss}) when expressing 
$\phi_a(x,p)$ as a function of space-time and momentum.
\be
\phi_a = - A_a \partial_{\rho} U^{\rho} +
C^a_{\mu\nu} \left( D^{\mu} U^{\nu} + D^{\nu} U^{\mu} 
+ \twoth \Delta^{\mu\nu} \partial_{\rho} U^{\rho} \right)
\label{phiexpand}
\ee
Here $A_a$ in general will depend on the scalar $U_{\alpha} p^{\alpha}$.  The 
tensor $C^a_{\mu\nu}$ could in principle be a linear combination of $g_{\mu\nu}$ 
and $p_{\mu}p_{\nu}$.  However, the former gives zero contribution.  Therefore 
we write $C^a_{\mu\nu} = C_a p_{\mu}p_{\nu}$ where $C_a$ will in general depend 
on the scalar $U_{\alpha} p^{\alpha}$.

It is worthwhile emphasizing that the expansion of $\Delta T^{\mu\nu}$ and 
$\phi_a$ in terms of the first order derivatives of the flow velocity is only an 
approximation.  It is referred to as the first order dissipative fluid dynamics.  
Inclusion of second order derivatives goes under the names of M\"uller and 
Israel and Stewart.  The second order theory is under intense investigation due 
to its usefulness in describing high energy nuclear collisions where space-time 
gradients are not necessarily small.  It is also worth emphasizing that these 
same quantities are zero for an equilibrated system in uniform flow.

It is now a straightforward matter to equate the two expressions for the 
dissipative part of the energy-momentum tensor.  It is advantageous to work in 
the local rest frame of the fluid.
\be
\zeta = \frac{1}{3} \sum_a \int \frac{d^3p}{(2\pi)^3} \frac{|{\bf p}|^2}{E_a} 
f_a^{\rm eq}(E_a/T) A_a(E_a)
\label{gasbulk}
\ee
\be
\eta = \frac{2}{15} \sum_a \int \frac{d^3p}{(2\pi)^3} \frac{|{\bf p}|^4}{E_a} 
f_a^{\rm eq}(E_a/T) C_a(E_a)
\label{gasshear}
\ee

How do we determine the $A_a$ and $C_a$?  The idea is to use the Boltzmann 
equation where the term $\partial f_a/\partial t + {\bf v}_a \cdot \nabla f_a$ 
is evaluated using the local equilibrium distribution $f_a^{\rm eq}(U_{\alpha} 
p^{\alpha}/T)$.  This is nonzero whenever the flow velocity is changing in 
space-time.  It will act as a source for the collision term on the other side of 
the Bolztmann equation.  With classical statistics the Boltzmann equation reads 
as follows.
\bd
E_a^{-1}p_a^{\mu} \partial_{\mu} f_a^{\rm eq} = f_a^{\rm eq}
\sum_{bcd} \frac{1}{1 + \delta_{ab}} \int \frac{d^3p_b}{(2\pi)^3} 
\frac{d^3p_c}{(2\pi)^3} \frac{d^3p_d}{(2\pi)^3} 
f_b^{\rm eq} W(a,b|c,d)
\ed
\bd
\times \left\{ \phi_c + \phi_d - \phi_a - \phi_b \right\}
+ f_a^{\rm eq} \sum_{cd} \int \frac{d^3p_c}{(2\pi)^3} \frac{d^3p_d}{(2\pi)^3}
W(a|c,d) \left\{ \phi_c +\phi_d -\phi_a\right\}
\ed
\be
+ \sum_{bc} \int \frac{d^3p_b}{(2\pi)^3} \frac{d^3p_c}{(2\pi)^3}
f_c^{\rm eq} \, W(c|a,b) \left\{ \phi_c -\phi_a -\phi_b  \right\}
\ee

The first task is to compute the left-hand side of the Boltzmann equation.  With 
$f_a^{\rm eq} = \exp(-U_{\nu}p^{\nu}/T)$ we have
\be\partial_{\mu} f_a^{\rm eq} = - \frac{1}{T} f_a^{\rm eq} p^{\nu} 
\left( \partial_{\mu} U_{\nu} - \frac{1}{T} U_{\nu} \partial_{\mu}T \right) \, .
\ee
Using the conservation equations for energy and momentum, 
$\partial_{\nu} T^{\mu\nu} = 0$, and entropy (since the viscous terms are 
neglected at this order), $\partial_{\mu} s^{\mu} = 0$, we may deduce that
\be
\Delta^{\mu\nu} \frac{1}{T} \partial_{\nu}T = - 
U^{\alpha}\partial_{\alpha}U^{\mu}
\label{dT}
\ee
which has solution
\be
\frac{1}{T} \partial_{\mu}T = U^{\alpha}\partial_{\alpha}U_{\mu}
+ \xi U_{\mu} \partial_{\alpha} U^{\alpha} 
\ee
where $\xi$ is a function of $T$ which is undetermined by Eq. (\ref{dT}).  
It can be determined by substituting the above expression into the 
conservation equations.  This gives $\xi = -s/c_V = -v_s^2$, where 
$c_V = d\epsilon/dT = Tds/dT$ is the heat capacity per unit volume 
and $v_s^2 = dP/d\epsilon$ is the square of the sound velocity.  Thus
\bd
p^{\mu} \partial_{\mu} f_a^{\rm eq} = - \frac{1}{T} f_a^{\rm eq} 
p^{\mu} p^{\nu} \left[ \partial_{\mu} U_{\nu} - 
U_{\nu} \left(U^{\alpha}\partial_{\alpha}\right) U_{\mu} \right]
\ed
\bd
= - \frac{1}{2T} f_a^{\rm eq} p^{\mu} p^{\nu} 
\Bigg[ \left( D_{\mu} U_{\nu} + D_{\nu} U_{\mu} 
+ \twoth \Delta_{\mu\nu} \partial_{\rho} U^{\rho} \right)
 - \twoth \Delta_{\mu\nu} \partial_{\rho} U^{\rho}
\ed
\be
 + 2 v_s^2 U_{\mu} U_{\nu} \partial_{\rho} U^{\rho} \Bigg] \, .
\label{pdelf}
\ee

After substituting in the structure of the $\phi$'s and grouping terms we get
\be
{\cal A}^a \left( \partial_{\rho} U^{\rho} \right)-
{\cal C}^a_{\mu\nu} \left( D^{\mu} U^{\nu} + D^{\nu} U^{\mu} 
+ \twoth \Delta^{\mu\nu} \partial_{\rho} U^{\rho} \right) = 0
\ee
where
\bd
{\cal A}^a = \frac{1}{3E_a T} \left[ \left(p_a^{\alpha}U_{\alpha} \right)^2      
\left( 1 - 3v_s^2 \right) - m_a^2 \right]
\ed
\bd
+ \sum_{bcd} \frac{1}{1 + \delta_{ab}} \int \frac{d^3p_b}{(2\pi)^3} 
\frac{d^3p_c}{(2\pi)^3} \frac{d^3p_d}{(2\pi)^3} 
f_b^{\rm eq} \, W(a,b|c,d)
\left\{ A_c + A_d - A_a - A_b \right\}
\ed
\bd
+ \sum_{cd} \int \frac{d^3p_c}{(2\pi)^3} \frac{d^3p_d}{(2\pi)^3}
W(a|c,d) \left\{ A_c + A_d - A_a \right\}
\ed
\be
+ \sum_{bc} \int \frac{d^3p_b}{(2\pi)^3} \frac{d^3p_c}{(2\pi)^3}
f_b^{\rm eq} \, W(c|a,b) \left\{ A_c - A_a - A_b \right\}
\label{calA}
\ee
and
\bd
{\cal C}_a^{\mu\nu} = \frac{p_a^{\mu} p_a^{\nu}}{2E_a T}
+ \sum_{cd} \int \frac{d^3p_c}{(2\pi)^3} \frac{d^3p_d}{(2\pi)^3}
W(a|c,d) \left\{ C_c p_c^{\mu} p_c^{\nu} + C_d p_d^{\mu} p_d^{\nu}
 - C_a p_a^{\mu} p_a^{\nu} \right\}
\ed
\bd
+ \sum_{bc} \int \frac{d^3p_b}{(2\pi)^3} \frac{d^3p_c}{(2\pi)^3}
f_b^{\rm eq} \, W(c|a,b) \left\{ C_a p_a^{\mu} p_a^{\nu} + 
C_b p_b^{\mu} p_b^{\nu} - C_c p_c^{\mu} p_c^{\nu} \right\}
\ed
\bd
+ \sum_{bcd} \frac{1}{1 + \delta_{ab}} \int \frac{d^3p_b}{(2\pi)^3} 
\frac{d^3p_c}{(2\pi)^3} \frac{d^3p_d}{(2\pi)^3} 
f_b^{\rm eq} \, W(a,b|c,d)
\ed
\be
\times \left\{ C_c p_c^{\mu} p_c^{\nu} + C_d p_d^{\mu} p_d^{\nu}
 - C_a p_a^{\mu} p_a^{\nu} - C_b p_b^{\mu} p_b^{\nu} \right\} \, .
\label{calC}
\ee

The bulk viscosity is very small or zero in several limits.  The first 
is the conformal limit, which means that the theory has no dimensional 
parameters, such as mass or intrinsic energy scale.  Then $P \sim T^4$ 
and $v_s^2 = 1/3$, and so the first term on the right hand side of 
Eq. (\ref{calA}), the source term, vanishes and so do the $A_a$.  The 
second is the nonrelativistic limit of a single species of particle. 
Then $P \sim m^{3/2} T^{5/2} \exp(-m/T)$ and $v_s^2 = T/m$ (plus corrections 
of higher order in $T/m$).  Once again the first term on the right hand 
side of Eq. (\ref{calA}) vanishes (to lowest order in $T/m$) and so the 
bulk viscosity should be very small.  These arguments do not apply to the shear 
viscosity since the source term does not involve the 
equation of state.

\section{Landau-Lifshitz Condition}

The equation (\ref{calA}) does not have a unique solution as it stands.  For 
example, consider elastic scattering for just one type of particle.  Starting 
with one solution $A(E)$ we can generate an infinite number of other solutions 
by making the shift $A(E) \rightarrow A'(E) = A(E) - a - bE$, where $a$ and $b$ 
are arbitrary constants.  These constants are associated with particle 
conservation ($a$) and energy conservation ($b$).  This has been noted in the 
literature before.  It may be restated in more physical terms.  To return a 
system to kinetic and chemical equilibrium after a change in volume, one might 
either change the number of particles while keeping the average energy per 
particle fixed, or one might change the average energy per particle while 
keeping the total number of particles fixed.  Now it is apparent that this 
ambiguity is associated with the Landau Lifshitz condition (\ref{LLdef}), which 
is also sometimes called the condition of fit when solving (\ref{calA}).

Consider an arbitrary set of particle species and all possible reactions allowed 
by the symmetries.  Make the shift $A_a(E_a) \rightarrow A'_a(E_a) = A_a(E_a) - 
a_a - bE_a$.  The constant $b$ must be the same for all species of particle.  
The constants $a_a$ are just like chemical potentials; they satisfy the same 
relationships among themselves.  Since we are restricting our considerations to 
systems with zero net quantum numbers, such as electric charge and baryon number, 
it is obvious that the $a_a$ are all zero, just as all chemical potentials are 
zero.  The constant $b$ acts like an inverse temperature and is as yet 
undetermined.

Suppose that we have a particular solution $A^{\rm par}_a$ to (\ref{calA}); does 
it satisfy the Landau Lifshitz condition (\ref{LLdef})?   The general solution 
would be $A_a(E_a) = A^{\rm par}_a(E_a) - bE_a$.  Using Eq. (\ref{deltaTuv}) the 
Landau Lifshitz condition for the $A$ term is
\be
\sum_a \int \frac{d^3p}{(2\pi)^3} f_a^{\rm eq}(E_a/T) E_a
\left[ A^{\rm par}_a(E_a) - bE_a \right] = 0 \, .
\ee
Here it is useful to know the contributions to the pressure, energy density, 
entropy density and heat capacity from a single species of particle. 
\ba
P_a &=& T \int \frac{d^3p}{(2\pi)^3} f_a^{\rm eq}(E_a/T) \nonumber \\
\epsilon_a &=& \int \frac{d^3p}{(2\pi)^3} E_a f_a^{\rm eq}(E_a/T) \nonumber \\
s_a &=& \frac{1}{3T^2} \int \frac{d^3p}{(2\pi)^3} 
|{\bf p}|^2 f_a^{\rm eq}(E_a/T) \nonumber \\
c_{Va} &=& \frac{1}{T^2}\int \frac{d^3p}{(2\pi)^3} E_a^2 f_a^{\rm eq}(E_a/T) \, . 
\ea
Now the coefficient $b$ is determined in terms of integrals of the particular 
solutions.
\be
b = \frac{1}{T^2 c_V} \sum_a \int \frac{d^3p}{(2\pi)^3} 
f_a^{\rm eq}(E_a/T) E_a A^{\rm par}_a(E_a) \, .
\ee 
If the particular solutions already happen to satisfy the Landau Lifshitz 
condition, then $b=0$.  Substitution of $A_a(E_a) = A^{\rm par}_a(E_a) - bE_a$ 
into Eq. (\ref{gasbulk}), with $b$ as determined above, gives an expression for 
the bulk viscosity.
\be
\zeta = \frac{1}{3} \sum_a \int \frac{d^3p}{(2\pi)^3E_a} f_a^{\rm eq}(E_a/T) 
A^{\rm par}_a(E_a) \left( |{\bf p}|^2 - 3 v_s^2 E_a^2 \right) 
\label{gasbulkpar}
\ee
Notice that if the particular solutions happen to satisfy the Landau Lifshitz 
condition then Eq. (\ref{gasbulkpar}) reduces to Eq. (\ref{gasbulk}).

There is no ambiguity with the $C_a$ in the shear viscosity because of the 
tensorial structure of the integrand in Eq. (\ref{calC}).  Physically the reason 
has to do with the fact that shear viscosity is associated with the response to 
changes in shape at fixed volume whereas bulk viscosity is associated with the 
response to changes in volume at fixed shape.
   
\section{Relaxation Time Approximation}

Consider the Boltzmann equation (\ref{Boltzeq}).  Let us suppose that all 
species of particles for all values of momentum are in equilibrium except for 
species $a$ with momentum ${\bf p}_a$.  Replace all phase space distributions 
$f$ with their equilibrium values $f^{\rm eq}$ except for $f_a$, which we allow 
to be out of equilibrium by a small amount.  Thus we write $f_a = f_a^{\rm eq} + 
\delta f_a$.  This is the momentum-dependent relaxation time approximation.  We 
approximate the Boltzmann equation by
\be
\frac{\partial f_a({\bf x},t,{\bf p}_a)}{\partial t} + 
{\bf v}_a \cdot \nabla f_a({\bf x},t,{\bf p}_a) =
- \omega_a(E_a) \delta f_a({\bf x},t,{\bf p}_a)
\label{relaxBoltzeq}
\ee 
where
\bd
\omega_a(E_a) =
\sum_{bcd} \frac{1}{1 + \delta_{ab}} \int \frac{d^3p_b}{(2\pi)^3} 
\frac{d^3p_c}{(2\pi)^3} \frac{d^3p_d}{(2\pi)^3} W(a,b|c,d)
f_b^{\rm eq}
\ed
\be
+ \sum_{cd} \int \frac{d^3p_c}{(2\pi)^3} \frac{d^3p_d}{(2\pi)^3} W(a|c,d) 
+ \sum_{bc} \int \frac{d^3p_b}{(2\pi)^3} \frac{d^3p_c}{(2\pi)^3} W(c|a,b) 
f_b^{\rm eq}
\label{freqint}
\ee 
is the frequency of interaction.  The equilibration time is defined as
\be
\tau_a(E) = \omega_a^{-1}(E) \, .
\ee
The deviation $\delta f_a$ is related to the function $\phi_a$ defined in Eq. 
(\ref{departure}) by
\be
\delta f_a(x,p) = f_a^{\rm eq}(x,p) \phi(x,p) \, .
\ee
Therefore we can substitute Eqs. (\ref{phiexpand}) and (\ref{pdelf}) into Eq. 
(\ref{relaxBoltzeq}) to solve for the functions $A^{\rm par}_a$ and $C_a$, where 
$C_a^{\mu\nu} = C_a p_a^{\mu} p_a^{\nu}$.
\be
A^{\rm par}_a(E_a) = \frac{\tau_a(E_a)}{3TE_a} \left[ \left( 1 -
3v_s^2 \right) E_a^2 - m_a^2 \right]
\label{Arelax}
\ee\be
C_a(E_a) = \frac{\tau_a(E_a)}{2TE_a}
\label{Crelax}
\ee

The viscosities are now readily calculated using these results.  As usual, it is 
advantageous to work in the local rest frame of the fluid.
\be
\zeta = \frac{1}{9T} \sum_a \int \frac{d^3p}{(2\pi)^3}
\frac{\tau_a(E_a)}{E_a^2} 
\left[ \left( 1 - 3v_s^2 \right) E_a^2 - m_a^2 \right]^2
f_a^{\rm eq}(E_a/T) \, .
\label{bulkrelaxkin}
\ee
\be\eta = \frac{1}{15T} \sum_a \int \frac{d^3p}{(2\pi)^3} \frac{|{\bf 
p}|^4}{E_a^2} 
\tau_a(E_a) f_a^{\rm eq}(E_a/T)
\label{gasshearrelax}
\ee
In the relaxation time approximation one must calculate the momentum dependent 
relation time or rate from Eq. (\ref{freqint}) and then substitute into the 
above expressions and perform a one dimensional integration.  These are 
generalizations of the formulas given in \cite{Gavin:1985ph} to an arbitrary 
number of species of particles with energy-dependent relaxation times.

As a further approximation one may calculate a mean interaction frequency 
$\bar{\omega_a}$ and an associated mean relaxation time $\bar{\tau_a} = 
\bar{\omega_a}^{-1}$ via
\ba
\bar{\omega_a} &=& \frac{1}{n_a} \int \frac{d^3p}{(2\pi)^3} \omega_a(E_a)
f_a^{\rm eq}(E_a/T) \nonumber \\
n_a &=& \int \frac{d^3p}{(2\pi)^3} f_a^{\rm eq}(E_a/T)
\ea
However, there is really no need to make this approximation unless one only 
desires a rough order of magnitude estimate.  Depending on the dynamics, the 
relaxation time may be highly momentum dependent.  In either case it is clear 
that the particles with the longest relaxation time dominate the viscosities, 
since these particles can transport energy and momentum over greater distances 
before interacting.

Now comes a subtle point.  What if we were to include weak interactions in our 
considerations?  Clearly the relaxation times for weak interactions are orders 
of magnitude greater than the relaxation times for the strong interactions, so 
they would dominate the viscosities.  The answer is that one must evaluate the 
actual physical conditions to which the viscous fluid equations are to be 
applied.  For example, in high energy nuclear collisions the size of the system 
is on the order of 10 fm while the lifetime is of order 10 fm/c.  Any electrons, 
positrons or neutrinos that might be produced by this system will simply escape 
and not interact with any of the hadrons.  They cannot transport energy and 
momentum to another part of the system.  In addition, due to the weakness of the 
interaction very few of them will actually be produced.  Thus the weak 
interactions are irrelevant in this situation.  The environment in the early 
universe or supernovae will most likely require inclusion of the effects of the 
weak interactions since the length and time scales are so much greater. 

\section{Mean Field or Potential Effects}

Most hadronic models of hot matter involve temperature dependent mean fields, 
such as the Walecka model or the linear sigma model.  In the absence of chemical 
potentials there should be no condensation of vector or tensor fields, only 
scalar fields.  Then the single particle energies have the form $E_a = 
\sqrt{{\bf p}^2 + \bar{m}_a^2(T)}$ where $\bar{m}_a(T)$ is a temperature 
dependent effective mass that arises from the mean fields, in other words, 
potential energy effects.  This affects the bulk viscosity, but not the shear 
viscosity, in several ways.

The Boltzmann equation acquires an extra term on the left side of Eq. 
(\ref{Boltz1})
\be
\left( \frac{\partial}{\partial t} + \frac{{\bf p}}{E_a} \cdot \nabla_x 
- \nabla_x E_a \cdot \nabla_p \right) f_a({\bf x}, t, {\bf p}) = C[f]
\ee
where $C[f]$ represents the collision, formation and decay terms.  The extra 
term involves the force ${\bf F} = d{\bf p}/dt = - \nabla_x E$.  To calculate 
the viscosities the left side is evaluated with the local equilibrium 
distribution
\be
f_a^{\rm eq} = \exp\left[-U_{\alpha}(x)p_a^{\alpha}(x)/T(x)\right] \, .
\ee
Now $p_a^0 = E_a$ depends on $x$ because $m$ depends on $T$ which depends on $x$.  
This approach is not new but has been proven or justified many times in the past.  
See, for example, references \cite{Perry,Jeon}.

A straightforward calculation gives
\bd
\left( \frac{\partial}{\partial t} + \frac{{\bf p}}{E_a} \cdot \nabla_x 
- \nabla_x E_a \cdot \nabla_p \right) f_a^{\rm eq} = 
\ed
\be
- \frac{1}{E_a T} f_a^{\rm eq} \left[ p^{\mu} p^{\nu} 
\left( \partial_{\mu} U_{\nu} - \frac{1}{T} U_{\nu} \partial_{\mu}T \right)
+2\frac{d\bar{m}^2_a}{dT} U^{\alpha} \partial_{\alpha} T \right] \, .
\ee
The gradient of the temperature was determined previously.
\be
\frac{1}{T} \partial_{\mu}T = U^{\alpha}\partial_{\alpha}U_{\mu}
- v_s^2 U_{\mu} \partial_{\alpha} U^{\alpha} 
\ee
This allows us to write
\bd
\left( \frac{\partial}{\partial t} + \frac{{\bf p}}{E_a} \cdot \nabla_x 
- \nabla_x E_a \cdot \nabla_p \right) f_a^{\rm eq} =
\ed
\bd 
- \frac{1}{2E_a T} f_a^{\rm eq} \Bigg[ p^{\mu} p^{\nu} 
\left( D_{\mu} U_{\nu} + D_{\nu} U_{\mu} 
+ \twoth \Delta_{\mu\nu} \partial_{\rho} U^{\rho} \right)
\ed
\be
 - \twoth \partial_{\rho} U^{\rho} \left( (1-3v_s^2)E_a^2 - \bar{m}_a^2
+ 3 v_s^2 T^2 \frac{d\bar{m}^2_a}{dT^2} \right) \Bigg] \, .
\label{pdelf2}
\ee
Thus the term $(1-3v_s^2)E_a^2 - m_a^2 = {\bf p}^2 - 3 v_s^2 E_a^2$ in Eq. 
(\ref{calA}) gets replaced by $(1-3v_s^2)E_a^2 - \bar{m}_a^2 + 3 v_s^2 T^2 
d\bar{m}^2_a/dT^2 = {\bf p}^2 - 3 v_s^2 (E_a^2 - T^2 d\bar{m}^2_a/dT^2)$.

Taking into account the mean field effects modifies the particular solution in 
the relaxation time approximation.
\be
A^{\rm par}_a(E_a) = \frac{\tau_a(E_a)}{3TE_a} \left[ \left( 1 - 3v_s^2 \right) 
E_a^2 - \bar{m}_a^2 + 3 v_s^2 T^2 \frac{d\bar{m}^2_a}{dT^2} \right]
\label{ArelaxMF}
\ee 
However, the mean fields also affect the equation of state, the speed of sound, 
and the Landau-Lifshitz condition, so it is not so straightforward to deduce the 
bulk viscosity at this point.

\subsection{Model with no symmetry breaking}

Consider a model with $N$ scalar fields $\Phi_a$ that has no symmetry breaking, 
meaning that $\langle \Phi_a \rangle = 0$ for all $a$.  It has the effective 
Lagrangian\be
{\cal L}_{\rm eff} = \thalf \sum_a \left( \partial_{\mu} \Phi_a \right)^2
- U(\Phi_1,...,\Phi_N)
\label{Leff}
\ee
Assume that the potential is a polynomial in the fields to arbitrarily high 
order.  It represents localized multi-particle interactions.  The quasi-particle 
approach includes both mean fields and independent thermal fluctuations around 
the mean fields, all calculated in a thermodynamically self-consistent manner 
\cite{fraser}-\cite{Mocsy:2004ab}.  This is sometimes referred to as the Phi-
derivable approach, and sometimes as the summation of daisy and super-daisy 
diagrams.  If fermions are present they are integrated out and the effects of their interactions are subsumed in $U$.  In this approximation only even powers of the fields in the potential play a role.  The thermal average may be written as
\be
\langle U \rangle = \sum_{n_1 \cdot\cdot\cdot n_N}
U_{n_1 \cdot\cdot\cdot n_N} \langle \Phi_1^2 \rangle^{n_1} \cdot\cdot\cdot 
\langle \Phi_N^2 \rangle^{n_N}
\ee
where the $U_{n_1 \cdot\cdot\cdot n_N}$ are constants.  The effective masses are 
obtained from
\be
\bar{m}_a^2 = \left\langle \frac{\partial^2 U}{\partial \Phi_a^2} \right\rangle
\ee
where it is assumed that the system has been diagonalized in terms of normal 
modes such that
\be
\left\langle \frac{\partial^2 U}{\partial \Phi_a \partial \Phi_b} \right\rangle 
= \delta_{ab} \bar{m}_a^2
\ee
From the combinatorics
\be
\left\langle \frac{\partial^2 U}{\partial \Phi_a^2} \right\rangle =
\sum_{n_1 \cdot\cdot\cdot n_N} U_{n_1 \cdot\cdot\cdot n_N}
\frac{(2n_a) (2n_a-1) (2n_a-3)!!}{(2n_a-1)!! \,\, \langle \Phi_a^2 \rangle} 
\langle \Phi_1^2 \rangle^{n_1} \cdot\cdot\cdot \langle \Phi_N^2 \rangle^{n_N}
\ee
so that
\be
\bar{m}_a^2 = \frac{2}{\langle \Phi_a^2 \rangle}
\sum_{n_1 \cdot\cdot\cdot n_N} U_{n_1 \cdot\cdot\cdot n_N} n_a 
\langle \Phi_1^2 \rangle^{n_1} \cdot\cdot\cdot \langle \Phi_N^2 \rangle^{n_N}
\label{meanmass}
\ee

The equation of state is given by
\ba
P &=& P_0 - V \nonumber \\
\epsilon &=& \epsilon_0 + V
\ea
where the subscript 0 refers to the free particle form with effective masses 
$\bar{m}_a$.  The energy-momentum tensor is
\be
T^{\mu\nu} = T_0^{\mu\nu} + g^{\mu\nu} V\label{TwithV}
\ee
Note that $Ts = \epsilon + P = \epsilon_0 + P_0 = Ts_0$ where $s=dP/dT$ and 
$s_0=dP_0/dT$ so that the form of the entropy is unchanged.  This is a 
consequence of the assumption of independent particle motion between collisions 
in the mean field approximation.  The potential energy density is obtained from
\ba
V &=& \langle U \rangle - \thalf \sum_a \bar{m}_a^2 \langle \Phi_a \rangle^2
\nonumber \\
&=& \sum_{n_1 \cdot\cdot\cdot n_N} U_{n_1 \cdot\cdot\cdot  n_N}
\left[ 1 - (n_1 + \cdot\cdot\cdot + n_N) \right] 
\langle \Phi_1^2 \rangle^{n_1} \cdot\cdot\cdot \langle \Phi_N^2 \rangle^{n_N}
\ea
The entropy density is computed from the formulas
\be
\frac{dP}{dT} = \sum_a \frac{\partial P_{0a}}{\partial T}
+ \sum_a \frac{\partial P_{0a}}{\partial \bar{m}_a^2}
\frac{d\bar{m}_a^2}{dT} - \frac{dV}{dT}
\ee
\be
\frac{\partial P_{0a}}{\partial \bar{m}_a^2} =
- \thalf \langle \Phi_a^2 \rangle
\ee
\be
\langle \Phi_a^2 \rangle = \int \frac{d^3p}{(2\pi)^3} \frac{1}{E_a} 
f_a^{\rm eq}(E_a/T)
\ee
whereby it is readily shown that
\be
\frac{dV}{dT} = - \thalf \sum_a \langle \Phi_a^2 \rangle
\frac{d\bar{m}_a^2}{dT}
\ee
from which follows
\be
\frac{dP}{dT} = \sum_a s_{0a}
\ee
so that the model is thermodynamically consistent.  There are some additional 
interesting temperature derivatives that can be derived, such as
\be
\frac{dV}{dT} = \sum_a V_a \frac{d\langle \Phi_a^2 \rangle}{dT}
\ee
and 
\be
\frac{d\langle U \rangle}{dT} = \frac{1}{2} \sum_a  \bar{m}_a^2
\frac{d\langle \Phi_a^2 \rangle}{dT}
\ee 

The numbers $U_{n_1 \cdot\cdot\cdot n_N}$ have the interpretation of 
representing the interaction of $2(n_1 + \cdot\cdot\cdot +n_N)$ particles in the 
initial plus final states.  Taking account of the combinatorics, the vertex 
itself would be
\bd
\frac{U_{n_1 \cdot\cdot\cdot n_N}}{(2n_1-1)!! \cdot\cdot\cdot (2n_N-1)!!}
\ed
where it is assumed that all $n_a \neq 0$.  If an $n_a$ happened to be zero then 
the factor $(2n_a-1)$ is replaced by 1.

In Fermi liquid theory, functional variation of the energy density with respect 
to the distribution functions should yield the single particle energies 
\cite{baym1991landau}.  That relationship holds here too.  The kinetic part of 
the energy density is $\epsilon_0$ with the effective masses $\bar{m}_a$.  
Making a variation 
$\delta f_a$ also affects the mass.  Hence
\ba
\frac{\delta \epsilon}{\delta f_a} &=& E_a + \sum_b \int \frac{d^3p}{(2\pi)^3}
\frac{ f_b(E_b) }{2E_b} \frac{\partial \bar{m}^2_b}{\partial \langle \Phi_a^2
\rangle} \frac{\delta \langle \Phi_a^2 \rangle}{\delta f_a}
+ \frac{\delta V}{\delta f_a} \nonumber \\
&=& E_a + \frac{1}{2} \frac{\delta \langle \Phi_a^2 \rangle}{\delta f_a}
\sum_b \frac{\partial \bar{m}^2_b}{\partial \langle \Phi_a^2 \rangle}
\langle \Phi_b^2 \rangle + \frac{\delta V}{\delta f_a}
\label{Fermivar} 
\ea
Variation of the potential part is given by
\ba
\delta V &=& \sum_{n_1 \cdot\cdot\cdot n_N} U_{n_1 \cdot\cdot\cdot n_N}
\left[ 1 - (n_1 + \cdot\cdot\cdot + n_N) \right] 
\langle \Phi_1^2 \rangle^{n_1} \cdot\cdot\cdot \langle \Phi_N^2 \rangle^{n_N}
\nonumber \\
&\times& \left[ \frac{n_1}{\langle \Phi_1^2 \rangle} \delta \langle \Phi_1^2 
\rangle + \cdot\cdot\cdot +
\frac{n_N}{\langle \Phi_N^2 \rangle} \delta \langle \Phi_N^2 \rangle \right]
\ea
so that
\be
\frac{\delta V}{\delta f_a} = \frac{1}{\langle \Phi_a^2 \rangle}
\frac{\delta \langle \Phi_a^2 \rangle}{\delta f_a}
\sum_{n_1 \cdot\cdot\cdot n_N} U_{n_1 \cdot\cdot\cdot n_N}
\left[ 1 - (n_1 + \cdot\cdot\cdot + n_N) \right] n_a
\langle \Phi_1^2 \rangle^{n_1} \cdot\cdot\cdot \langle \Phi_N^2 \rangle^{n_N}
\ee
Now it is easy to see from the expression (\ref{meanmass}) for the mass that 
$\partial \bar{m}_a^2/\partial \langle \Phi_b^2 \rangle$ is a symmetric matrix 
and that
\be
\frac{\delta V}{\delta f_a} = - \frac{1}{2} \frac{\delta \langle \Phi_a^2 
\rangle}{\delta f_a} \sum_b \frac{\partial \bar{m}^2_b}{\partial \langle 
\Phi_a^2 \rangle} \langle \Phi_b^2 \rangle
\ee
This cancels the extra term from $\epsilon_0$ so that
\be
\frac{\delta \epsilon}{\delta f_a} = E_a
\ee
verifying the internal consistency of the model.

\subsection{Model with symmetry breaking}

Now we allow for one of the fields to condense.  For definiteness let it be the 
$N'th$ one.  After making the shift $\Phi_N \rightarrow \Phi_N + v$, where $v$ 
is the condensate, all fields obey $\langle \Phi_a \rangle = 0$.  Generalizing 
the previous analysis we write
\be
\langle U \rangle = \sum_{n_1 \cdot\cdot\cdot n_N} \sum_l
U_{n_1 \cdot\cdot\cdot n_N;l} \langle \Phi_1^2 \rangle^{n_1} \cdot\cdot\cdot 
\langle \Phi_N^2 \rangle^{n_N} v^l
\ee
If there is no condensation, then all $U_{n_1 \cdot\cdot\cdot n_N;l}$ with $l > 
0$ vanish, and we return to the previous case.  Following the same arguments as 
before we obtain the effective masses
\be
\bar{m}_a^2 = \frac{2}{\langle \Phi_a^2 \rangle}
\sum_{n_1 \cdot\cdot\cdot n_N} \sum_l U_{n_1 \cdot\cdot\cdot n_N;l} n_a 
\langle \Phi_1^2 \rangle^{n_1} \cdot\cdot\cdot 
\langle \Phi_N^2 \rangle^{n_N} v^l
\ee
and the potential energy density
\ba
V &=& \langle U \rangle - \thalf \sum_a \bar{m}_a^2 \langle \Phi_a \rangle^2
\nonumber \\
&=& \sum_{n_1 \cdot\cdot\cdot n_N} \sum_l U_{n_1 \cdot\cdot\cdot n_N;l}
\left[ 1 - (n_1 + \cdot\cdot\cdot + n_N) \right] 
\langle \Phi_1^2 \rangle^{n_1} \cdot\cdot\cdot 
\langle \Phi_N^2 \rangle^{n_N} v^l
\ea 
The value of the condensate is determined by extremizing the pressure $P=P_0-V$ 
at fixed $T$, namely $(\partial P/\partial v)_T=0$.  The result is
\be
\sum_{n_1 \cdot\cdot\cdot n_N} \sum_l
U_{n_1 \cdot\cdot\cdot n_N;l} l \langle \Phi_1^2 \rangle^{n_1} \cdot\cdot\cdot 
\langle \Phi_N^2 \rangle^{n_N} v^l = 0
\label{vequation}
\ee
This is just the same condition as $\langle \partial U/\partial v \rangle = 0$.

For the dissipative part of the energy-momentum tensor and the 
Landau-Lifshitz condition we need
\ba
\Delta V &=& \sum_{n_1 \cdot\cdot\cdot n_N} \sum_l
U_{n_1 \cdot\cdot\cdot n_N;l}
\left[ 1 - (n_1 + \cdot\cdot\cdot + n_N) \right] 
\langle \Phi_1^2 \rangle^{n_1} \cdot\cdot\cdot \langle \Phi_N^2 \rangle^{n_N}
v^l \nonumber \\
&\times& \left[ \frac{n_1}{\langle \Phi_1^2 \rangle} \Delta \langle \Phi_1^2 
\rangle + \cdot\cdot\cdot
\frac{n_N}{\langle \Phi_N^2 \rangle} \Delta \langle \Phi_N^2 \rangle \right]
\nonumber \\
&-& \Delta v \sum_{n_1 \cdot\cdot\cdot n_N} \sum_l
U_{n_1 \cdot\cdot\cdot n_N;l}
(n_1 + \cdot\cdot\cdot + n_N) l 
\langle \Phi_1^2 \rangle^{n_1} \cdot\cdot\cdot \langle \Phi_N^2 \rangle^{n_N}
v^{l-1} 
\ea
The condensate deviates from its equilibrium value because the thermal 
fluctuations deviate.  One may express $\Delta v$ in terms of the $\Delta 
\langle \Phi_a^2 \rangle$ using Eq. (\ref{vequation}) but for our purposes there 
is no need to do so explicitly.  For verification of the Fermi liquid result 
that functional variation of the energy density with respect to the distribution 
function yields the single particle energy, it is sufficient to observe that
\be
\frac{\delta V}{\delta f_a} = - \frac{1}{2} \frac{\delta \langle \Phi_a^2 
\rangle}{\delta f_a} \sum_b \frac{\partial \bar{m}^2_b}{\partial \langle 
\Phi_a^2 \rangle} \langle \Phi_b^2 \rangle
- \frac{1}{2} \frac{\delta v}{\delta f_a} \sum_b \frac{\partial 
\bar{m}^2_b}{\partial v} \langle \Phi_b^2 \rangle 
\ee
This exactly cancels the extra term coming from $\epsilon_0$ so that
\be
\frac{\delta \epsilon}{\delta f_a} = E_a
\ee
in the presence of a condensate too.  From this point on, the expression for 
$\Delta T^{\mu\nu}$ and the bulk and shear viscosities is the same as when there 
is no condensation.

\subsection{Landau-Lifshitz condition}

The single particle energy at finite temperature is a functional of the 
distribution functions $E_a  = E_a \left(\left\{f\right\}\right)$. When there 
is a small deviation from equilibrium 
\be
\label{deq1}
f_a \left (x, p\right) = f_a^{\rm eq}\left(E_{a,0}\right) +
\delta f_a \left(x, p\right) \,,  
\ee
the single particle energy changes to 
\be
E_a = E_{a,0} + \delta E_a \,. \label{deq1_1}
\ee
Here the subscript $0$ indicates the value the energy would have if there was no 
departure from equilibrium at all.  If $f_a^{\rm eq}$ is expressed as a function 
of the true energy $E_a$, then  
\be
f_a \left (x, p\right) = f_a^{\rm eq}\left(E_a\right) +\delta \tilde{f_a}
 \left(x, p\right) \, , 
\ee
where
\be 
\delta \tilde{f_a} \left(x, p\right)  = \delta f_a \left(x, p\right) - 
\frac{\partial f_a^{\rm eq}\left(E_a\right)}
{\partial E_a} \delta E_a \, .
\label{deq2}  \ee
The structure of the quasi-particle Boltzmann equation is such that the 
it is the function $\delta \tilde{f_a}$ which determines the transport 
coefficients.  It is important to realize that energy conservation in the 
collisions dictates that the linearization of the collision term has to be 
performed with respect to the true energy $E_a$. 

Following the same arguments as earlier, expansion of $T^{ij}$ from 
Eq. (\ref{TwithV}) around local equilibrium, using Eq. (\ref{deq1}), 
leads to 
\be
\label{delTij}
\Delta T^{ij} = \sum_a  \int \frac{d^3p}{\left(2\pi\right)^3}
\frac{p^i p^j}{E_a} \delta \tilde{f}_a \,. 
\ee 
The change in the energy density is given by 

\ba
\label{llcondition}\Delta T^{00} &=& \sum_a  \int 
\frac{d^3p}{\left(2\pi\right)^3} E_a \delta f_a 
\nonumber \\
&=& \sum_a  \int \frac{d^3p}{\left(2\pi\right)^3}\left( E_a \delta \tilde{f_a}  
+ E_a \frac{\partial f_a^{\rm eq}\left(E_a\right)}{\partial E_a} 
 \delta E_a \right) \nonumber \\
&=& \sum_a  \int \frac{d^3p}{\left(2\pi\right)^3}\left( E_a \delta \tilde{f_a}  
- \frac{E_a}{T} f_a^{\rm eq}\left(E_a\right) \delta E_a \right) \nonumber \\
& = &  \sum_a  \int \frac{d^3p}{\left(2\pi\right)^3}\left( E_a \delta 
\tilde{f_a} - \frac{1}{2T}   \frac{d\bar{m}_a^2}{dT} 
f_a^{\rm eq}\left(E_a\right) \delta T\right) \, .
\ea    
Recall that
\be
\delta f_a = - {\rm e}^{-E_a/T} \left[\frac{\delta E_a}{T} - 
\frac{E_a}{T^2}\delta T\right]
\ee
and
\be
\delta \tilde{f_a} = {\rm e}^{-E_a/T} \frac{\delta T}{T^2} \, .
\ee  
Substituting these in (\ref{llcondition}) we find that, in the local rest frame,  
\be
\Delta T^{00} = \sum_a  \int \frac{d^3p}{\left(2\pi\right)^3} \frac{1}{E_a}
\left(E_a^2 - \bar{m}_a T \frac{d\bar{m}_a}{dT}\right)\delta \tilde{f_a} \,.
\ee
Hence, in a general frame of reference, 
\begin{equation}
\Delta T^{\mu\nu} = \sum_a  \int \frac{d^3p}{\left(2\pi\right)^3}
\frac{1}{E_a}\left(p_a^\mu p_a^\nu - U^\mu U^\nu 
\frac{d\bar{m}_a^2}{dT^2}\right)\delta \tilde{f_a} \, .
\end{equation}
This is the obvious generalization of the result of \cite{Jeon} to a system with 
multiple species of particles.  

Following the usual arguments we can deduce the viscosities.
\be
\zeta = \frac{1}{3} \sum_a \int \frac{d^3p}{(2\pi)^3 E_a} f_a^{\rm eq}(E_a/T) 
A_a(E_a) |{\bf p}|^2
\label{bulkMF} 
\ee
\be
\eta = \frac{2}{15} \sum_a \int \frac{d^3p}{(2\pi)^3 E_a} 
f_a^{\rm eq}(E_a/T) C_a(E_a) \, |{\bf p}|^4
\label{shearMF}
\ee

If one has a particular solution that does not happen to satisfy the Landau-
Lifshitz condition, one can make it so by adding a term linear in the 
energy, just as before.
\be
\sum_a \int \frac{d^3p}{(2\pi)^3E_a} f_a^{\rm eq}(E_a/T) 
\left[ E_a^2 - T^2 \frac{d \bar{m}_a^2}{dT^2} \right]
\left[ A^{\rm par}_a(E_a) - bE_a \right] = 0 \, .
\ee
To simplify the resulting formula, it is helpful to use
\be
\sum_a \int \frac{d^3p}{(2\pi)^3}
f_a^{\rm eq}(E_a/T) \left[ |{\bf p}|^2 - 3 v_s^2 
\left( E_a^2-T^2 \frac{d\bar{m}_a^2}{dT^2} \right) \right] = 0
\ee 
which is a consequence of the identity $dP/dT = (dP/d\epsilon) d\epsilon/dT = 
v_s^2 d\epsilon/dT$ in the independent particle models used here.  The 
coefficient $b$ is thus
\be
b = \frac{v_s^2}{T^2 s}
\sum_a \int \frac{d^3p}{(2\pi)^3E_a} f_a^{\rm eq}(E_a/T) 
\left[ E_a^2 - T^2 \frac{d \bar{m}_a^2}{dT^2} \right] A^{\rm par}_a(E_a)
\ee
Substitution of $A_a(E_a) = A^{\rm par}_a(E_a) - bE_a$ into Eq. (\ref{bulkMF}) 
gives
\be
\zeta = \frac{1}{3} \sum_a \int \frac{d^3p}{(2\pi)^3 E_a} f_a^{\rm eq}(E_a/T) 
A^{\rm par}_a(E_a) \left[ |{\bf p}|^2 - 3 v_s^2 
\left( E_a^2-T^2 \frac{d\bar{m}_a^2}{dT^2} \right) \right]
\label{bulkMFb} 
\ee
Of course, the term proportional to $v_s^2$ will integrate to zero if the 
particular solution does satisfy the Landau-Lifshitz condition.  The appearance 
of the factor in square brackets is natural since it matches the source function 
in Eq. (\ref{calA}).

In the relaxation time approximation this becomes
\be
\zeta = \frac{1}{9T} \sum_a \int \frac{d^3p}{(2\pi)^3}
\frac{\tau_a(E_a)}{E_a^2} f_a^{\rm eq}(E_a/T)
\left[ |{\bf p}|^2 - 3 v_s^2 
\left( E_a^2-T^2 \frac{d\bar{m}_a^2}{dT^2} \right) \right]^2 
\, .
\label{bulkrelaxmf}
\ee
This expression is obviously positive definite.
\section{The Linear $\sigma$ Model}
 
The linear $\sigma$ model has long been used as a simple renormalizable model of 
pion dynamics at low energy.  Although it cannot claim to be quantitatively 
accurate, being supplanted by chiral perturbation theory, it is still a 
much-used model for testing approximations and as a proxy for more involved and 
detailed low energy models of QCD.  Indeed, we will show that the model 
exemplifies the richness of the equation of state and transport coefficients to 
be expected from QCD.

The Lagrangian is 
\be
\label{langrangian_sigma_model}
{\cal L} = \frac{1}{2}\left(\partial_\mu \sigma \right)^2 + 
\frac{1}{2}\left(\partial_\mu \mbox{\boldmath $\pi$}\right)^2 
 - U\left(\sigma, \mbox{\boldmath $\pi$}\right)\,, 
\ee
where 
\be
\label{u_pi_sigma}
U\left(\sigma, \mbox{\boldmath $\pi$}\right) = \frac{\lambda}{4}
\left(\sigma^2 +\mbox{\boldmath $\pi$}^2 
   - f^2 \right)^2 - H\sigma \,.
\ee
Compare to Eq. (\ref{Leff}), where $\pi_i$ is identified with $\Phi_i$ for 
$i=1,2,3$ and $\sigma$ is identified with $\Phi_4$.  The $SU(2)_R \times 
SU(2)_L$ chiral symmetry is  explicitly broken by the term $H\sigma$ which gives 
the pion a mass. The scalar field has a non-vanishing vacuum expectation value 
$v$ determined at the classical level by the equation 
\begin{equation}
\label{equation_for_condensate}
\lambda v\left(v^2 - f^2\right) = H \,. 
\end{equation} The scalar field is thus split into a condensate and a 
fluctuation, 
$\sigma = v + \Delta$.  
The three parameters, $\lambda$, $H$ and $f$, are determined by the vacuum 
values of the pion decay constant $f_\pi$ and the pion and sigma masses.
\be
\lambda = \frac{m^2_{\sigma} - m_{\pi}^2}{2f_\pi^2} 
\ee
\be
 H =f_\pi^2 m_{\pi}^2
\ee
\be
f^2 = \frac{m_{\sigma}^2 - 3m_{\pi}^2}{m_{\sigma}^2 - m_{\pi}^2}f_\pi^2    
\ee
For numerical calculations we take $f_\pi = 93$ MeV, $m_{\pi} = 140$ MeV, and 
either $m_{\sigma} = 600$ or $900$ MeV. 

\subsection{Thermodynamics} 

The equation of state is calculated following the general procedure outlined in 
section 6.  It must be done numerically.  For more details the reader is 
referred to \cite{fraser}-\cite{Mocsy:2004ab}.

The temperature dependence of the meson masses and $\sigma$ condensate $v$ are 
shown in Fig. 1 for two illustrative values of the vacuum $\sigma$ mass. There 
is no true phase transition, only a crossover from the low temperature regime 
where there is a large difference between the two masses and the high 
temperature regime where the masses are practically identical.  The condensate 
decreases to very small values at high temperature but never vanishes.  The 
temperature at which the symmetry is approximately restored is around 245 MeV. 
It should be noted for future reference that the pion and $\sigma$ masses rise 
linearly with temperature at high temperature.  This is a generic feature of 
high temperature field theories.

\begin{figure}[!htbp]
\begin{center}
\includegraphics[width=4.5in,angle=0]{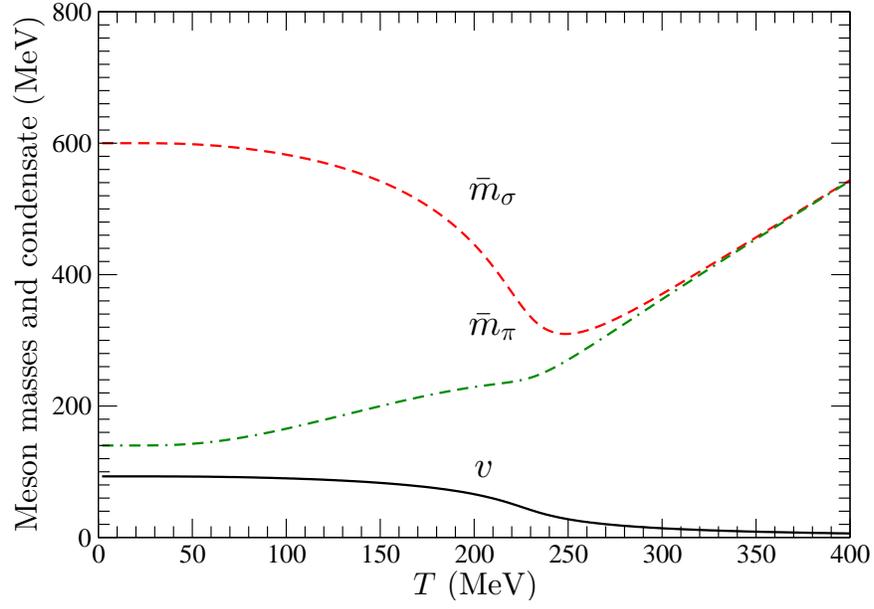}
\end{center}
\vspace{0.15in}
\begin{center}
\includegraphics[width=4.5in,angle=0]{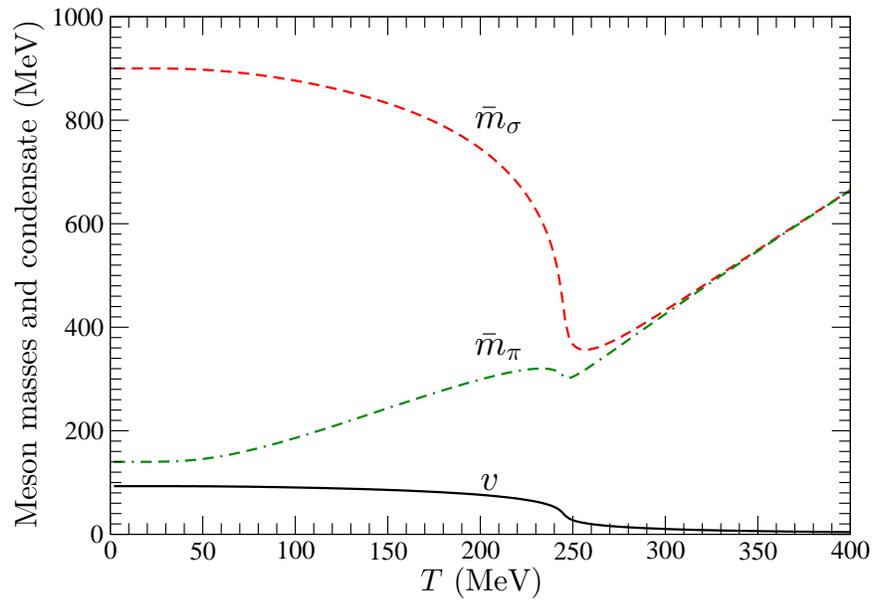}
\caption{(Color online) Variation of meson masses and condensate with temperature for a vacuum sigma mass of 600 MeV (top) and 900 MeV (bottom).}
\end{center}
\label{mass_and_condensate_vs_temp}
\end{figure}

The entropy density, energy density and pressure are shown in Fig. 2. These 
thermodynamic quantities follow a continuous curve with no phase transition. 
They go to zero exponentially as $T \rightarrow 0$ because all mesons are 
massive.  At large temperature they show the usual behavior that $s \sim T^3$ 
and $P \sim \epsilon \sim T^4$.  At intermediate temperatures rapid variation of 
the meson masses result in maxima in $s/T^3$ and $\epsilon/T^4$ but not in 
$P/T^4$.
 
\begin{figure}[!htbp]
\begin{center}
\includegraphics[width=4.3in,angle=0]{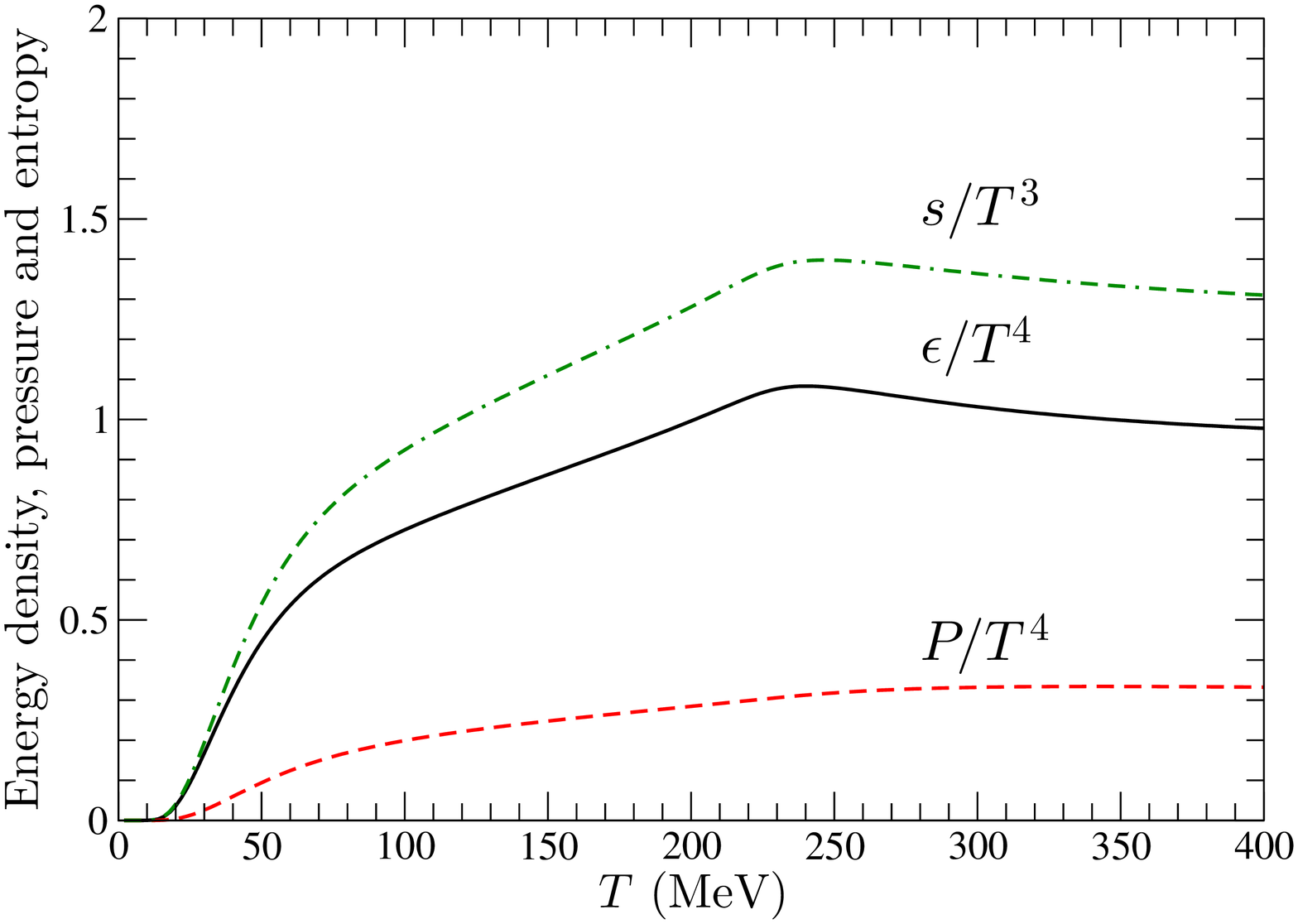}
\end{center}
\vspace{0.15in}
\begin{center}
\includegraphics[width=4.3in,angle=0]{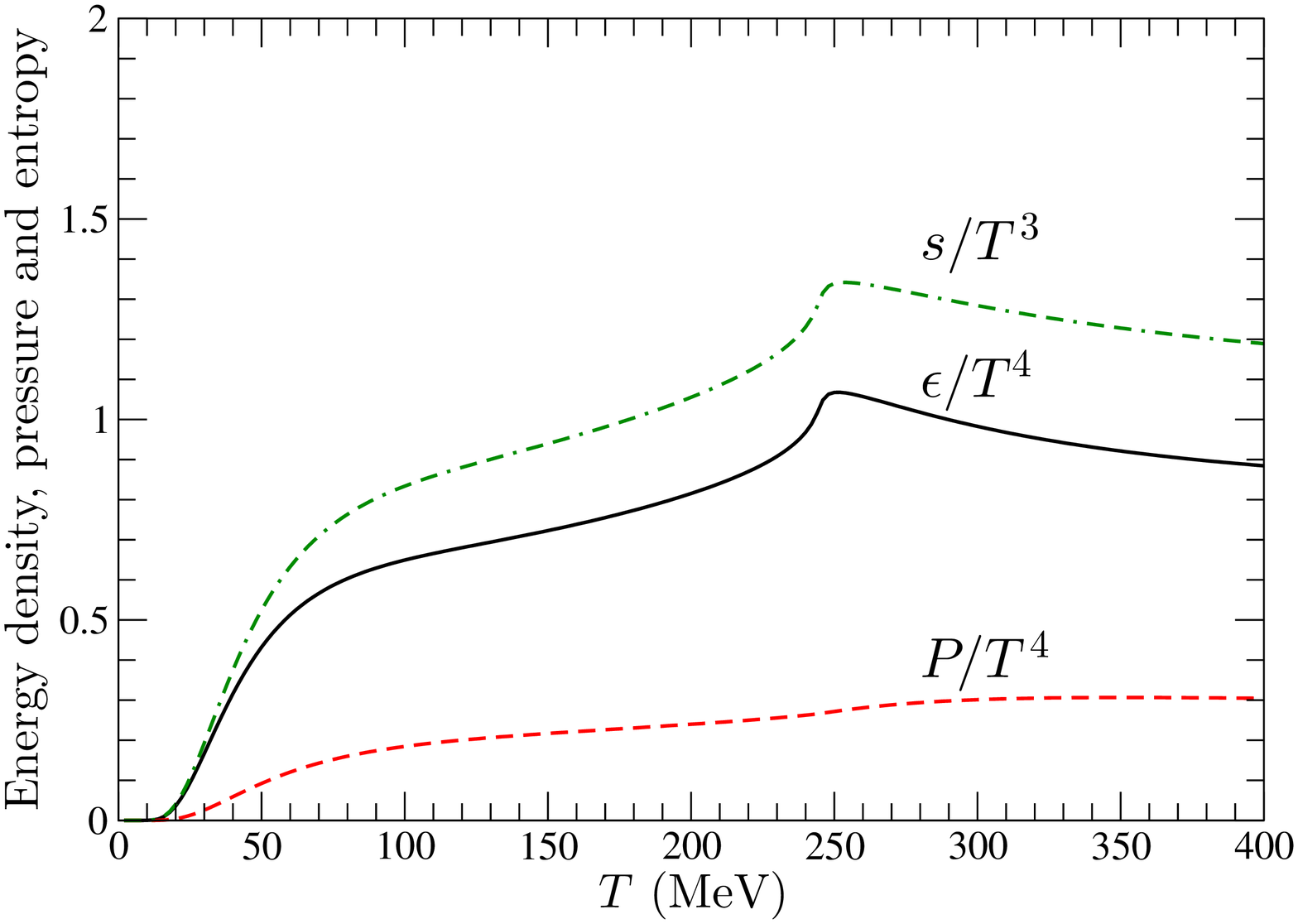}
\caption{(Color online) Energy density, pressure and entropy density as functions of temperature for a vacuum sigma mass of 600 MeV (top) and 900 MeV (bottom).}
\end{center}
\label{energy_pressure_entropy_figs}
\end{figure}

Figure 3 shows the speed of sound squared $v_s^2$ and heat capacity $c_V$ for 
the two choices of the vacuum $\sigma$ mass. As the vacuum $\sigma$ mass 
increases the speed of sound develops a dip and the heat capacity develops a 
peak around 245 MeV.  This indicates that the system is near a second order 
phase transition.  At high temperature $v_s^2 \rightarrow 1/3$ and $c_V 
\rightarrow T^3$, both on account of the fact that the equation of state 
approaches $P \sim T^4$. 
  
\begin{figure}[!htbp]
\begin{center}
\includegraphics[width=4.3in,angle=0]{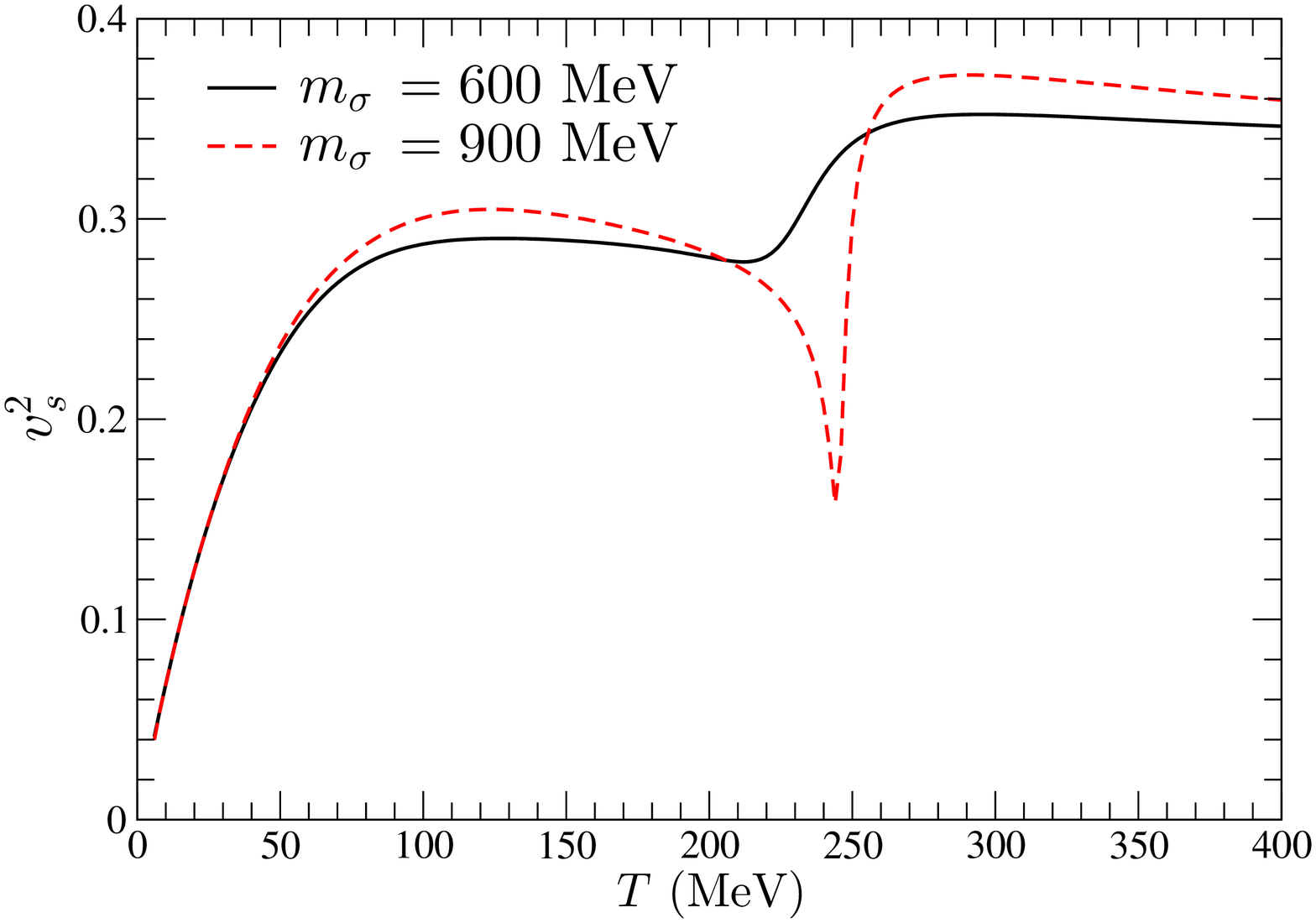}
\end{center}
\vspace{0.2in}
\begin{center}
\includegraphics[width=4.3in,angle=0]{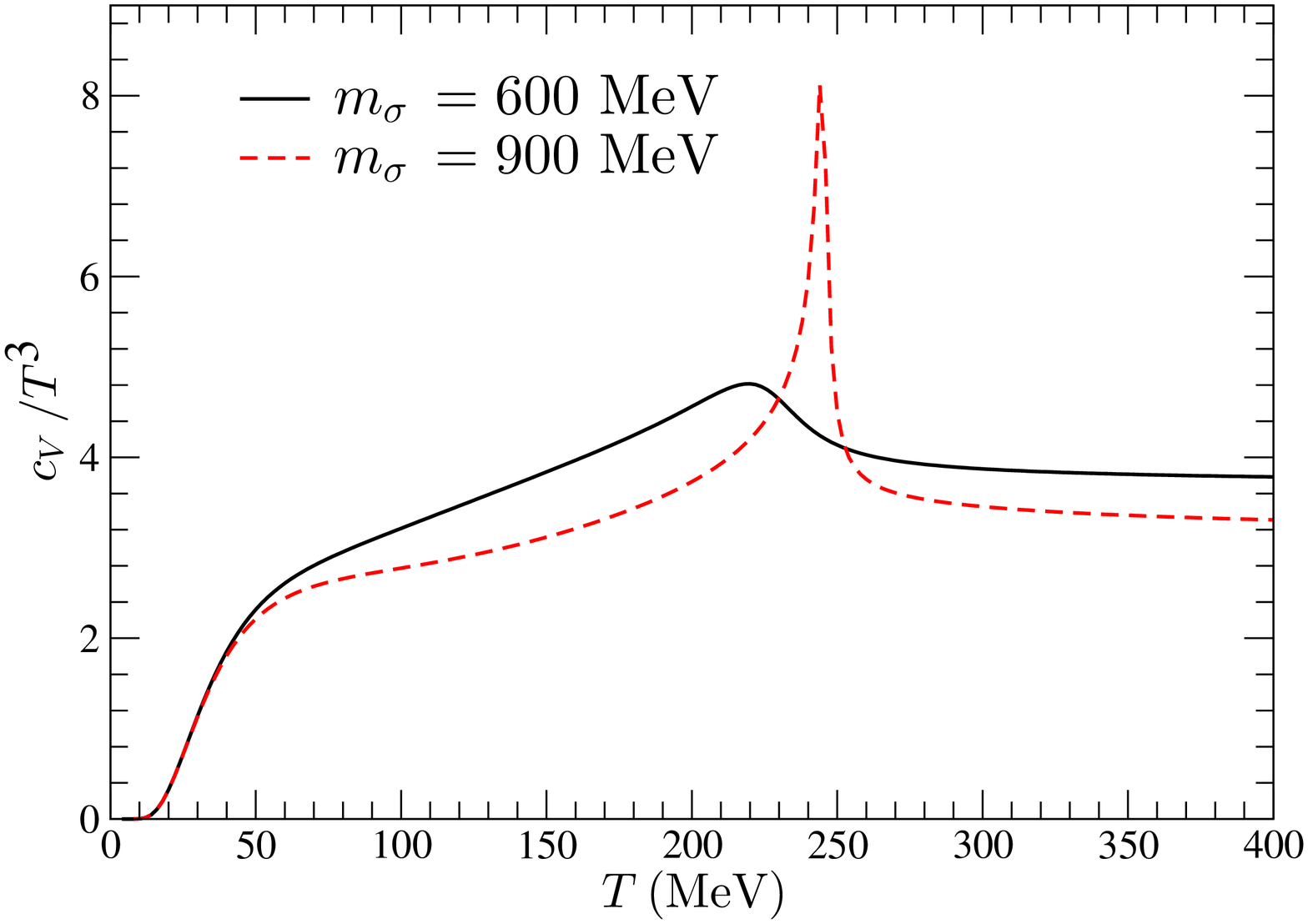}
\caption{(Color online) Temperature dependence of the speed of sound (top) and heat capacity (bottom).}
\end{center}
\label{speed_of_sound_and_sp_heat_figs}
\end{figure}

\subsection{Scattering amplitudes}
 
The Feynman rules can easily be determined from the Lagrangian of the linear 
$\sigma$ model.  At one loop order the various scattering amplitudes are as 
follows:   
\be
\label{amplitude_pi_pi_to_pi_pi1}
{\cal M}_{\pi^a\pi^b;\pi^c\pi^d} 
= -2 \lambda \left(
\frac{s - m_\pi^2}{s - m_\sigma^2}\delta_{ab}\delta_{cd} +
\frac{t - m_\pi^2}{t - m_\sigma^2}\delta_{ac}\delta_{bd} + 
\frac{u - m_\pi^2}{u - m_\sigma^2}\delta_{ad}\delta_{bc}\right) \,,
\ee
\be
{\cal M}_{\pi\sigma;\pi\sigma} = 
-2 \lambda -4\lambda^2 f_\pi^2 \left(\frac{3}{t - m_\sigma^2} +
\frac{1}{u - m_\pi^2} + \frac{1}{s - m_\pi^2}\right) \,,
\ee
\be
\label{amplitude_pi_pi_to_sigma_sigma}
{\cal M}_{\pi\pi;\sigma\sigma} = -2 \lambda
- 4\lambda^2 f_\pi^2 \left(\frac{3}{s - m_\sigma^2} +
\frac{1}{t - m_\pi^2} + \frac{1}{u - m_\pi^2}\right) \,,
\ee 
\be
\label{amplitude_sigma_sigma_to_sigm_sigma}
{\cal M}_{\sigma\sigma;\sigma\sigma} = -6 \lambda
- 36\lambda^2 f_\pi^2 \left(\frac{1}{s - m_\sigma^2} +
\frac{1}{t - m_\sigma^2} + \frac{1}{u - m_\sigma^2}\right) \,.
\ee
The poles in the $s$ and $u$ channels of $\pi \pi \to \pi \pi$ and $\pi \sigma 
\to \pi \sigma$, respectively, pose trouble as is well known in the literature. 
We are not aware of any prescription to regularize both of these singularities 
while satisfying crossing symmetry. It is obvious that the problematic terms are 
coming from the 3-point vertices. Those terms were not included when calculating 
the equation of state anyway. To be consistent with the equation of state, these 
terms must be dropped when calculating the viscosities.  Equivalently we 
approximate the scattering amplitudes by their limits as $s$, $t$ and $u$ all go 
to infinity. Thus the scattering amplitudes just reduce to constants.       

\subsection{Viscosities}
 
With the transition amplitudes specified, the departure functions $A_a$ and 
$C_a$ for $a = \pi$ and $\sigma$ can be determined as solutions to the integral 
equations ${\cal A}_a = 0$ and ${\cal C}_a^{\mu\nu} = 0$ using (\ref{calA}) and 
(\ref{calC}).  When substituted into (\ref{bulkMF}) and (\ref{shearMF}) we will 
get the bulk and shear viscosities.  

The set of integral equations are projected into a suitable vector space and 
solved by a variational method \cite{degroot,ferziger_kaper}. The departure 
functions are expanded in terms of some chosen basis functions $g_n$ as
\be
C_a = \sum_{n=1}^N c_{a,n} g_n\left(E_a/T\right) \,,
\ee   
and the expansion coefficients $c_{a,n}$ are adjusted to minimize the functional 
\begin{equation}
{\cal F} = \sum_a \left(C_a - \tilde{C}_a, C_a - \tilde{C}_a\right)\,,  
\end{equation}
where $\tilde{C}$ is the true solution.  We have defined the inner product between two functions $\chi_1$ and $\chi_2$ as
\begin{equation}
\left(\chi_1,\chi_2\right) = \sum_a \int\,\frac{d^3p}{\left(2\pi\right)^3} 
\chi_{1,a}\left(\mathbf{p}\right) 
{\cal R}\left[\chi_{2,a}\left(\mathbf{p}\right)\right] \,.
\end{equation}
where ${\cal R}$ is the collision operator acting on the space of departure
functions which, in the notation of (\ref{boltz_gen}), should be understood as  
\be
{\cal R}\left[\chi\right] =
\sum_{\{i\}\{j\}} \frac{1}{S} \int^{\prime} dP_i \, dP_j
W(\{i\}|\{j\})  \prod_{i=1}^n \prod_{j=1}^m f_i
\left(1 + \left(-1\right)^{s_j}f_j\right)  \left(\sum_i \chi_i 
- \sum_j \chi_j\right)\,.
\ee
With the variational technique one hopes to find a good approximation to the 
departure functions using a modest set of basis functions. However, this is not 
guaranteed.  In this paper we will show numerical results obtained this way for 
the shear viscosity.  Solving the integral equations for the bulk viscosity is 
more difficult, as is well known, due to the presence of zero modes.  Details of 
the numerical techniques used and their convergence properties will be presented 
elsewhere. 

\begin{figure}[!htbp]
\begin{center}
\includegraphics[width=4.3in,angle=0]{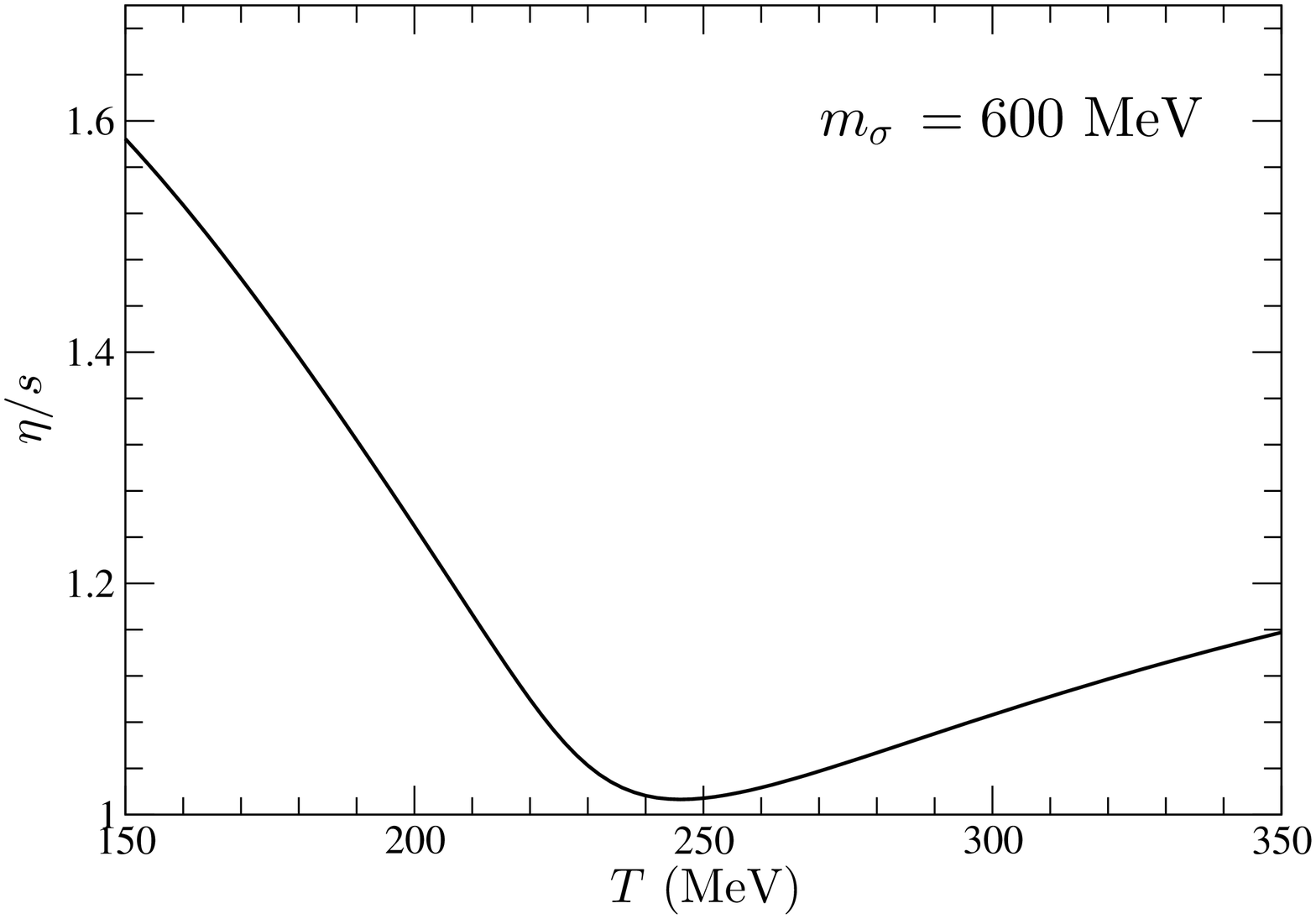}
\end{center}
\vspace{0.2in}
\begin{center}
\includegraphics[width=4.3in,angle=0]{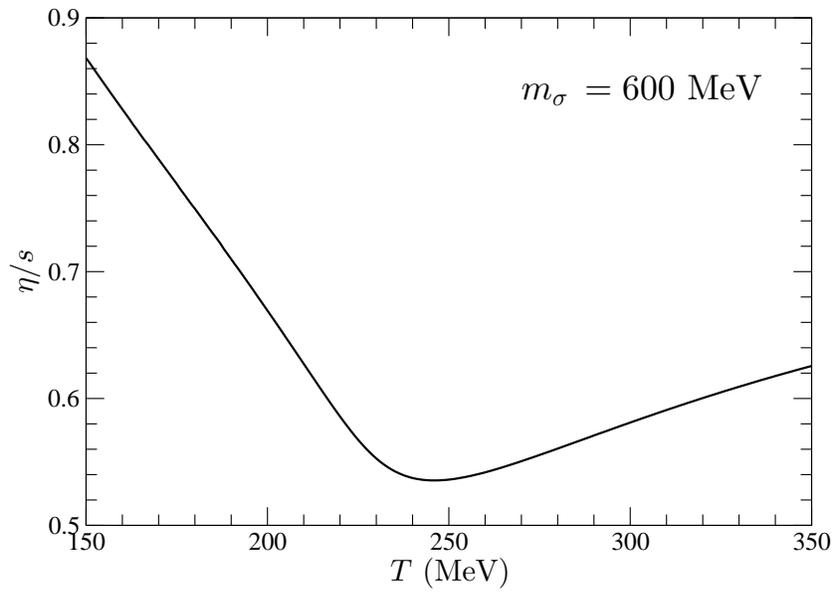}
\caption{Ratio of shear viscosity to entropy density in the Chapman-Enskog calculation (top) and in the relaxation time approximation (bottom). The vacuum sigma mass is 600 MeV.}
\end{center}
\label{shear_viscosity_600}
\end{figure}

\begin{figure}[!htbp]
\begin{center}
\includegraphics[width=4.3in,angle=0]{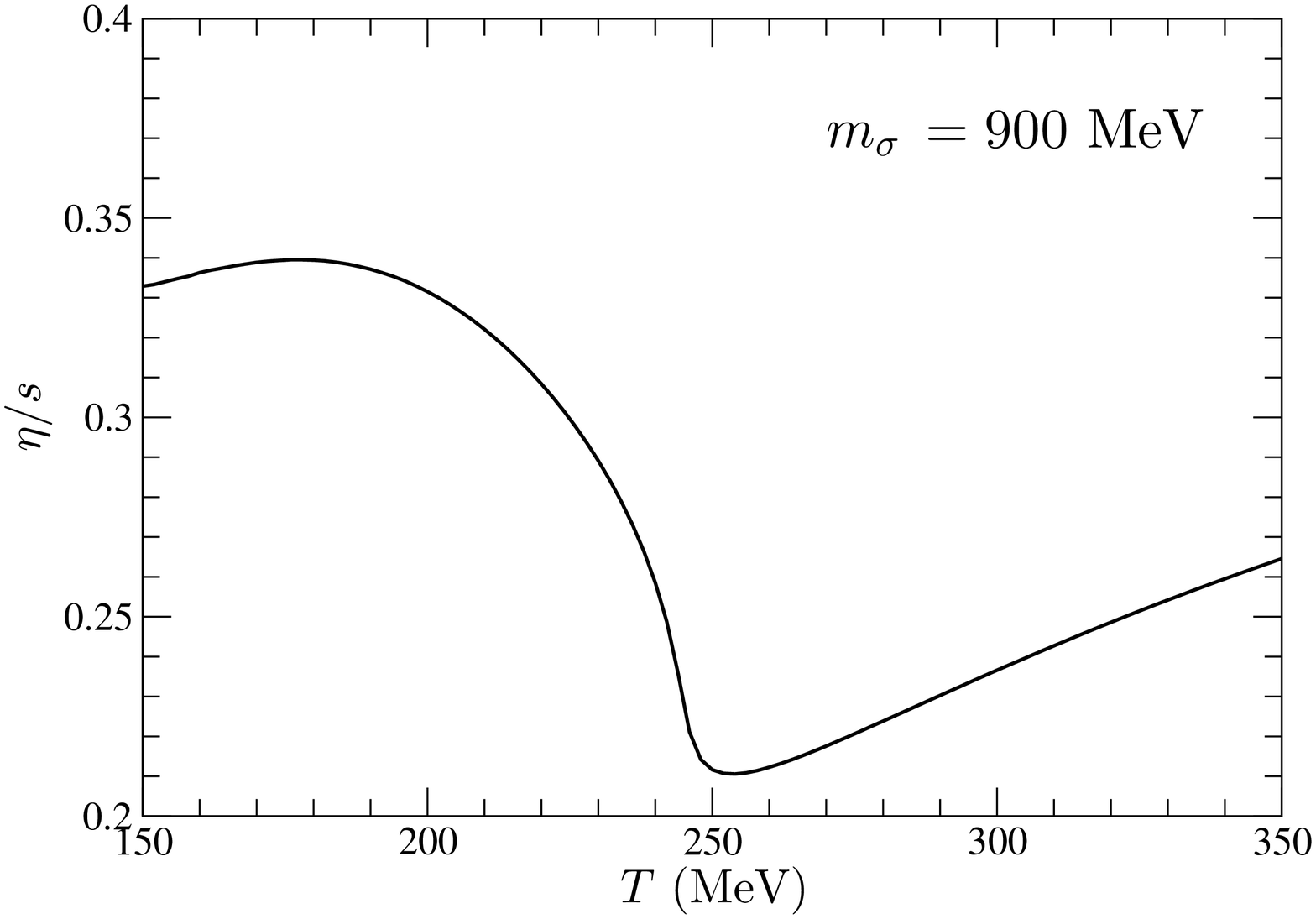}
\end{center}
\vspace{0.2in}
\begin{center}
\includegraphics[width=4.3in,angle=0]{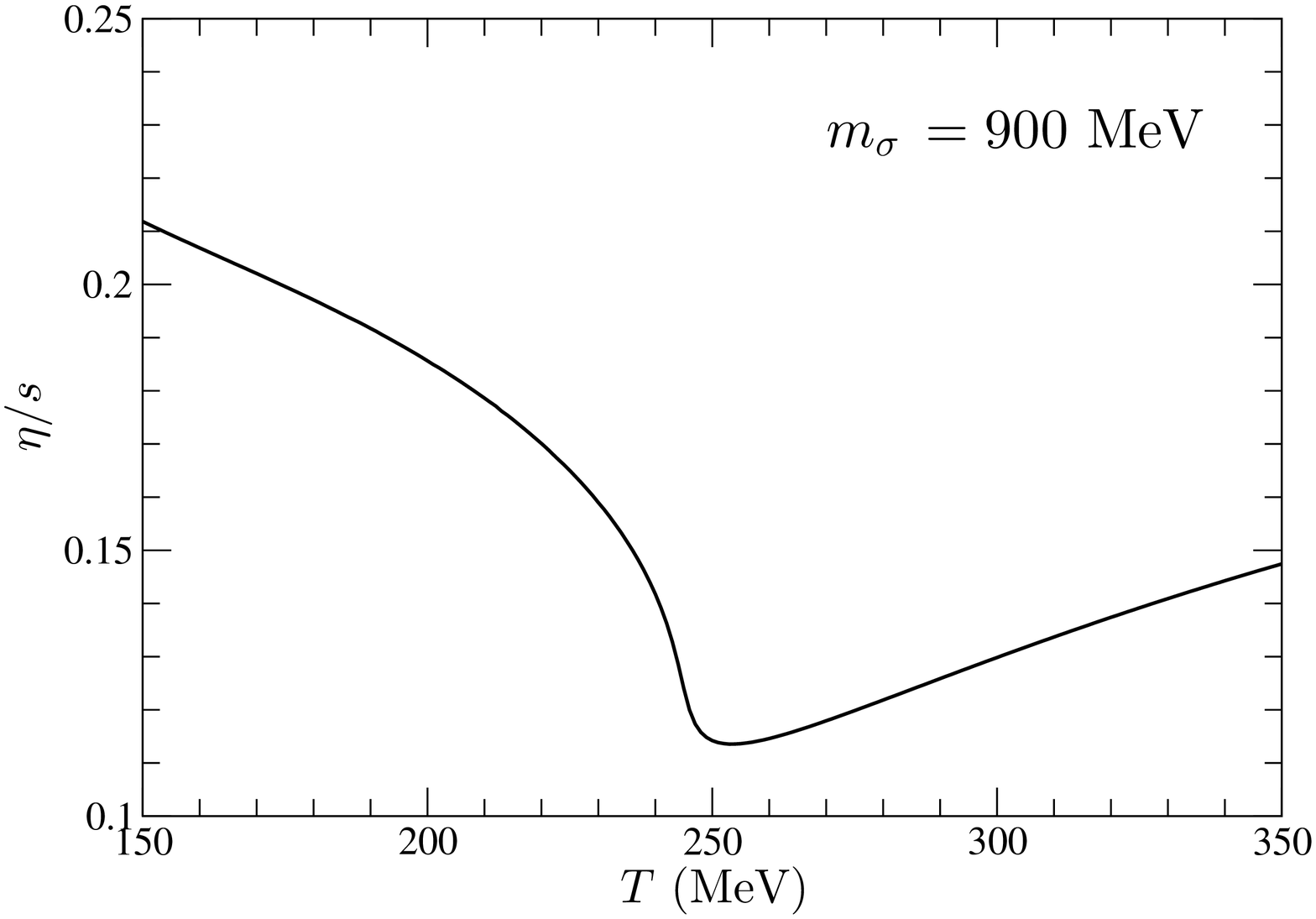}
\caption{Ratio of shear viscosity to entropy density in the Chapman-Enskog calculation (top) and in the relaxation time approximation (bottom). The vacuum sigma mass is 900 MeV.}
\end{center}
\label{shear_viscosity_900}
\end{figure}

In Figs. 4 ($m_{\sigma}$ = 600 MeV) and 5 ($m_{\sigma}$ = 900 MeV) we show the 
ratio of shear viscosity to entropy density from a 3rd order variational 
calculation compared to the relaxation time method. The 
variational method converges very quickly. For the present case the first order 
solution was already good to within a few percent and it was unnecessary to go 
beyond the 3rd order.  It is satisfying that results from the relaxation time 
approximation are within a factor of 2 compared to the sophisticated variational 
method. A minimum near the crossover temperature of 245 MeV is seen in both 
cases as expected.  The difference between the results of the variational method 
versus the relaxation time approximation is most apparent at low temperature and 
for $m_\sigma = 900$ MeV.  This can be attributed to how the collision dynamics 
is approximated when using the relaxation time method.  Although the relaxation 
time used is energy dependent, it is still only an approximation to the full 
solution of the integral equations.    

\newpage

\begin{figure}[!htbp]
\begin{center}
\includegraphics[width=4.3in]{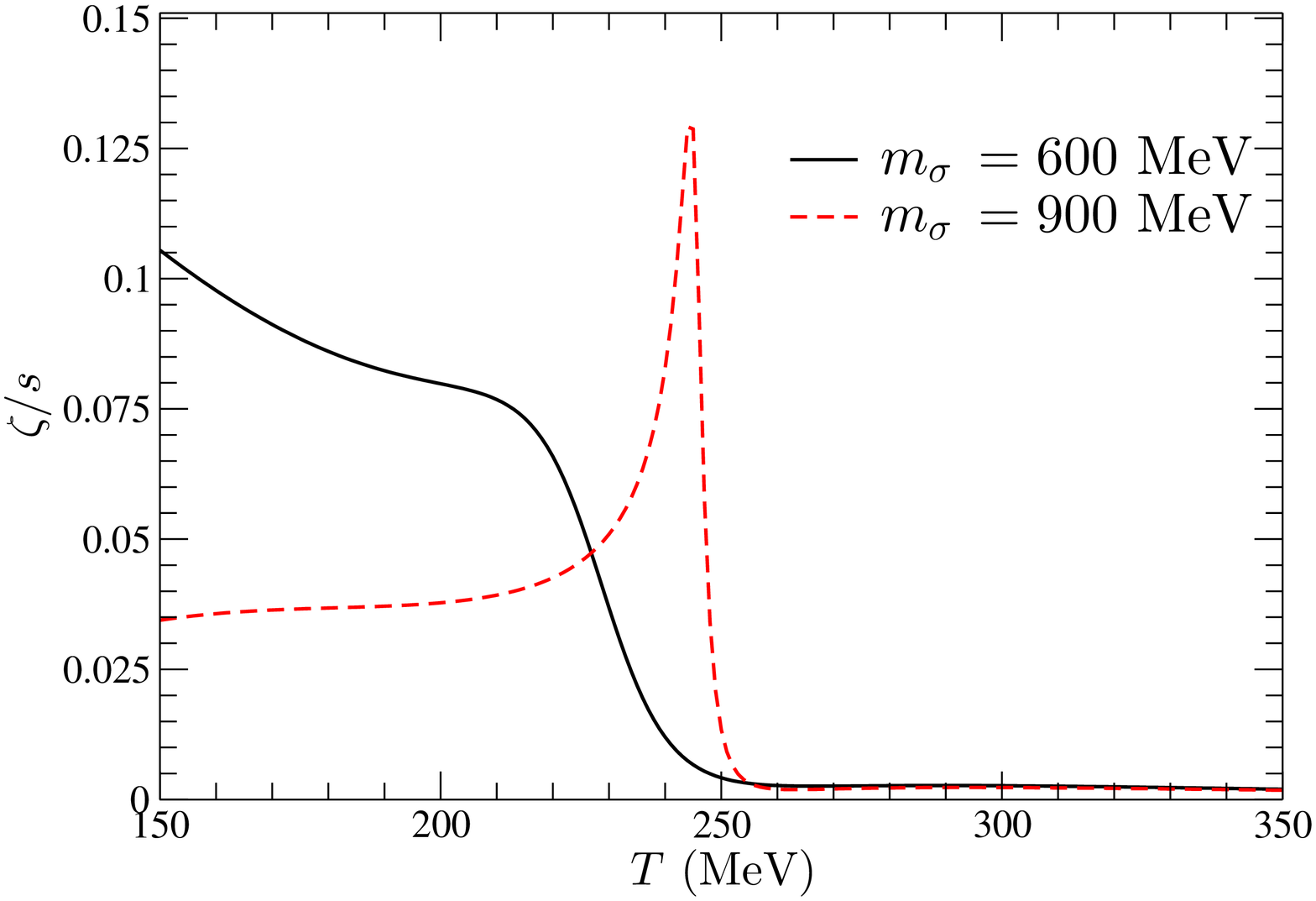}
\end{center}
\caption{(Color online) Ratio of bulk viscosity to entropy density in the relaxation time approximation.}
\label{zeta_only}
\end{figure}

The ratio of bulk viscosity to entropy density from the relaxation time 
approximation
\begin{eqnarray}
\zeta &=&  \frac{1}{T} \sum_a \int \frac{d^3p}{(2\pi)^3}
\frac{\tau_a(E_a)}{E_a^2} f_a^{\rm eq}(E_a/T)\nonumber \\
&\times& 
\left[\left(\frac{1}{3} - v_s^2\right) \left|\mathbf{p}\right|^2
 - v_s^2\left(\bar{m}_a^2 - T^2\frac{d\bar{m}_a^2}{dT^2}\right)\right]^2
 \,, 
\label{bulk4fig}
\end{eqnarray}
is shown in Fig. 6. Although we have not yet managed to solve the integral equations for the bulk viscosity, we do not expect the difference between the two approaches to be significantly different from what we obtained for the shear viscosity.  For a vacuum $\sigma$ mass of 900 MeV, there is a clear maximum in $\zeta/s$ near the crossover temperature.  To ascertain its origin it is only 
necessary to examine Eq. (\ref{bulk4fig}).  The ratio is quadratically proportional to the violation of conformality, either in terms of speed of sound
\bd
\oneth - v_s^2
\ed
or in terms of bulk transport mass
\bd
\bar{m}_a^2 - T^2\frac{d\bar{m}_a^2}{dT^2} =
\frac{d}{d\beta^2} \left( \beta^2 \bar{m}_a^2 \right) \,. 
\ed
A comparison of Figs. 3 and 7 with Fig. 6 clearly shows that $\zeta/s$ is 
largest when the violation of conformality is greatest.  
Note that for the system to be conformal it is only necessary that $P \sim T^4$ and that $\bar{m} \sim T$; it is not necessary for the effective mass to vanish.  Intuitively, the bulk viscosity is enhanced when it is easy to transfer energy between kinetic motion and internal degrees of freedom, such as resonances or heavier mass particle, mean fields, particle production or absorption, or effective masses that vary strongly with temperature.  All of these are playing a role here to a greater or lesser extent. 
\begin{figure}[!htbp]
\begin{center}
\includegraphics[width=4.3in,angle=0]{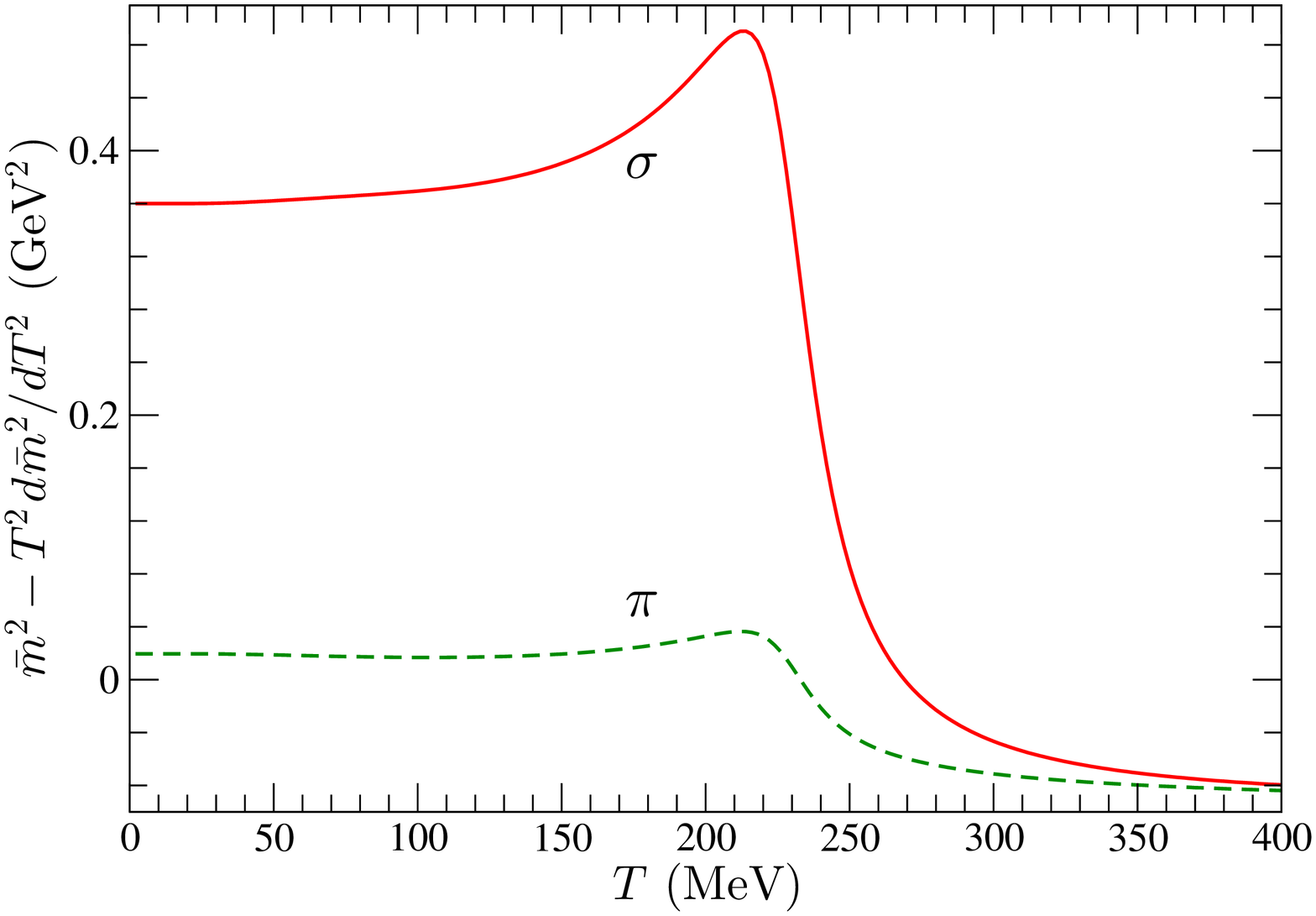}
\end{center}
\vspace{0.2in}
\begin{center}
\includegraphics[width=4.3in,angle=0]{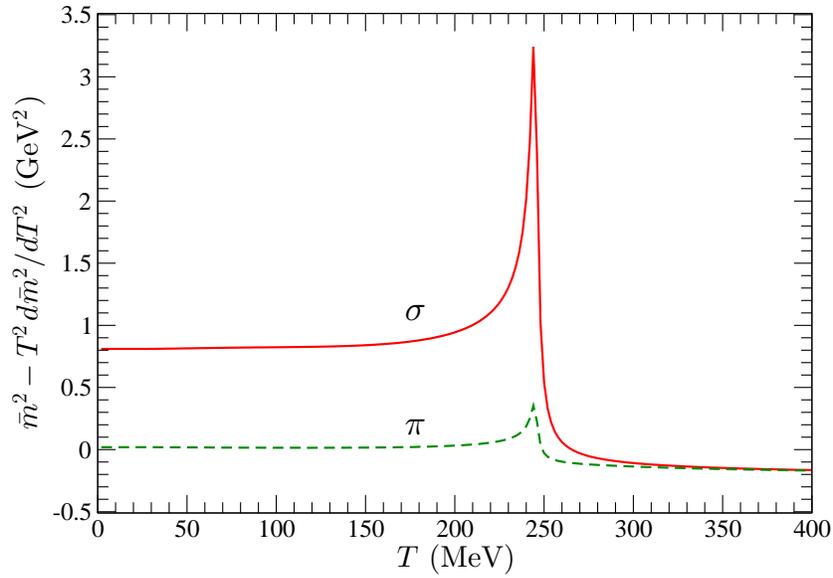}
\caption{(Color online) The quantity $\bar{m}_a^2 – d\bar{m}_a^2/dT^2$ which enters the calculation of the bulk viscosity for a vacuum sigma mass of 600 MeV (top) and 900 MeV (bottom).}
\end{center}
\label{mass_delta_figs}
\end{figure}

\section{Conclusion}

In this paper we constructed a theoretical framework for the calculation of the 
viscosities of hot hadronic matter.  The framework has the benefit that it is 
relativistic, it allows for an arbitrary number of hadron species, it allows for 
elastic and inelastic collisions and the formation and decay of resonances, it 
respects detailed balance, and it allows for temperature-dependent mean fields 
and temperature-dependent masses.  It is a consistent theory in the sense that 
the same interactions that are used to calculate the equation of state are used 
to calculate the viscosities.  The significance of this is that the bulk 
viscosity, in particular, depends very dramatically on the equation of state and 
on the quasi-particle masses.  The essential assumption is that quasi-particles 
are good degrees of freedom.  The basic resulting formulas have been assembled 
in the appendix for ease of reference.

As a nontrivial application we applied this theoretical framework to the linear 
$\sigma$ model using the physical vacuum pion mass.  As the vacuum $\sigma$ mass 
increases, the crossover transition become sharper, the speed of sound develops 
a dip, and the heat capacity develops a peak.  This is well-known physics, but 
it has significance for the viscosities.  As expected, the ratio $\eta/s$ has a 
minimum at the crossover temperature while the ratio $\zeta/s$ develops a 
maximum there for large enough vacuum $\sigma$ masses.

As a byproduct of our theoretical framework we generalized the relaxation time 
formulas for the shear and bulk viscosities of Gavin \cite{Gavin:1985ph} to 
include an arbitrary number of species of hadrons with energy-dependent relation 
times and temperature-dependent effective masses.  These formulas alone can be 
applied very usefully without much numerical work.

The theoretical framework presented here can be applied to increasingly 
sophisticated models of hadronic interactions.  Work on doing so is in progress.  
Also in progress is the extension to include chemical potentials.  The latter 
may not be important at RHIC and LHC but they will be for heavy ion collisions 
at FAIR.

\section*{Acknowledgements}

We thank M. Prakash, S. Gavin and S. Jeon for providing valuable input. P. C. 
thanks A. Hager and L. Keek for computational support, and E. Frodermann and E. 
S. Bowman for helpful discussions.  This work was supported by the US Department 
of Energy (DOE) under grant DE-FG02-87ER40328.

\section*{Appendix}

In this appendix we gather some of the basic results of the paper.  For more 
details see the text.

The Boltzmann equation is
\be
\left( \frac{\partial}{\partial t} + \frac{{\bf p}}{E_a} \cdot \nabla_x 
- \nabla_x E_a \cdot \nabla_p \right) f_a({\bf x}, t, {\bf p}) =
\sum_{\{i\}\{j\}} \frac{1}{S} \int^{\prime} dP_i \, dP_j
W(\{i\}|\{j\}) F\left[f\right] \, ,
\ee
where the prime indicates that there is no integration over the momentum of $a$.  
There is a statistical factor for identical particles in the initial state
\be
S = \prod_i n_i!
\ee
and products of Bose-Einstein and Fermi-Dirac distributions as appropriate
\be  
F\left[ f \right] = \prod_i \prod_j \left\{ f_j  
\left(1 + \left(-1\right)^{s_i}f_i\right) -  f_i
\left(1 + \left(-1\right)^{s_j}f_j\right) \right\} \,.
\ee
The rate function is
\be
W(\{i\}|\{j\}) = \frac{(2\pi)^4 \delta^4(P_i - P_j)}
{\left( \displaystyle{\prod_i 2E_i} \right) 
\left( \displaystyle{\prod_j 2E_j} \right)}
|{\cal M}(\{i\}|\{j\})|^2 \,.
\ee
The departure of the phase space distributions are expressed in terms of 
$\phi_a$ as
\be
f_a = f_a^{\rm eq} \left(1+\phi_a \right) \, ,
\ee
which furthermore has the tensorial decompostion
\be
\phi_a = - A_a \partial_{\rho} U^{\rho} +
C^a_{\mu\nu} \left( D^{\mu} U^{\nu} + D^{\nu} U^{\mu} 
+ \twoth \Delta^{\mu\nu} \partial_{\rho} U^{\rho} \right) \, .
\ee
In terms of these the viscosities are
\be
\zeta = \frac{1}{3} \sum_a \int \frac{d^3p}{(2\pi)^3} \frac{|{\bf p}|^2}{E_a} 
f_a^{\rm eq}(E_a/T) A_a(E_a) \,,
\ee
\be
\eta = \frac{2}{15} \sum_a \int \frac{d^3p}{(2\pi)^3} \frac{|{\bf p}|^4}{E_a} 
f_a^{\rm eq}(E_a/T) C_a(E_a) \,.
\ee
The departure functions satisfy integral equations.  For example, for two body 
reactions
\bd
\frac{1}{E_a T} \left[\left(\frac{1}{3} - v_s^2\right) \left|\mathbf{p}\right|^2 
- v_s^2\left(\bar{m}_a^2 - T^2\frac{d\bar{m}_a^2}{dT^2}\right) \right] =
\ed
\be
\sum_{bcd} \frac{1}{1 + \delta_{ab}} \int \frac{d^3p_b}{(2\pi)^3} 
\frac{d^3p_c}{(2\pi)^3} \frac{d^3p_d}{(2\pi)^3} 
f_b^{\rm eq} \, W(a,b|c,d)
\left\{ A_a + A_b - A_c - A_d \right\} \,,
\ee
and
\bd
\frac{p_a^{\mu} p_a^{\nu}}{2E_a T}
= \sum_{bcd} \frac{1}{1 + \delta_{ab}} \int \frac{d^3p_b}{(2\pi)^3} 
\frac{d^3p_c}{(2\pi)^3} \frac{d^3p_d}{(2\pi)^3} 
f_b^{\rm eq} \, W(a,b|c,d)
\ed
\be
\times \left\{ C_a p_a^{\mu} p_a^{\nu} + C_b p_b^{\mu} p_b^{\nu}
 - C_c p_c^{\mu} p_c^{\nu} - C_d p_d^{\mu} p_d^{\nu} \right\} \,.
\ee
For resonance formation and decay and multiparticle reactions see the text.
The dissipative part of the energy-momentum tensor can be written as
\be
\Delta T^{\mu\nu} = \sum_a  \int \frac{d^3p}{\left(2\pi\right)^3}
\frac{1}{E_a}\left(p_a^\mu p_a^\nu - U^\mu U^\nu 
\frac{d\bar{m}_a^2}{dT^2}\right)\delta \tilde{f_a}
\ee
where $\delta \tilde{f_a}$ is defined in the text.
In the relaxation time approximation the viscosities can be written as
\be
\eta = \frac{1}{15T} \sum_a \int \frac{d^3p}{(2\pi)^3} \frac{|{\bf 
p}|^4}{E_a^2} 
\tau_a(E_a) f_a^{\rm eq}(E_a/T)
\ee
and
\be
\zeta = \frac{1}{T} \sum_a \int \frac{d^3p}{(2\pi)^3}
\frac{\tau_a(E_a)}{E_a^2} f_a^{\rm eq}(E_a/T)
\left[ \left(\frac{1}{3} - v_s^2\right) \left|\mathbf{p}\right|^2 
- v_s^2\left(\bar{m}_a^2 - T^2\frac{d\bar{m}_a^2}{dT^2}\right) \right]^2 \,.
\ee
The effective masses are calculated self-consistently using the same potential 
$U$ and same approximations as when calculating the equation of state, namely
\be
\bar{m}_a^2 = \left\langle \frac{\partial^2 U}{\partial \Phi_a^2} \right\rangle 
\, .
\ee

\end{document}